\documentclass[
 reprint,
superscriptaddress,
 amsmath,amssymb,
 aip, apl
]{revtex4-2}

\usepackage{pifont}
\usepackage{graphicx}% Include figure files
\usepackage{bm}% bold math
\usepackage{sidecap}
\usepackage{xcolor}
\usepackage[colorlinks=true,citecolor=blue]{hyperref}
\usepackage{natbib}

\makeatletter

\newcommand{\Rmnum}[1]{\expandafter\@slowromancap\romannumeral #1@}
\makeatother

\begin{document}

\title{Terahertz Polarization Conversion from Optical Dichroism in a Topological Dirac Semimetal}% Force line breaks 

\author{Haiyu Meng}

\affiliation{School of Physics and Optoelectronics, Xiangtan University, Xiangtan 411100, China}
%\affiliation{Science, Mathematics and Technology (SMT), Singapore University of Technology and Design, Singapore 497372}
\affiliation{Department of Physics, National University of Singapore, Singapore 117542}
%\affiliation{Center for High-Resolution Electron Microscopy, College of Materials Science and Engineering, Hunan University, Changsha 410082, China}

\author{Lingling Wang}
\affiliation{School of Physics and Electronics, Hunan University, Changsha 410082, China}

\author{Ching Hua Lee}
\thanks{Authors to whom correspondence should be addressed: phylch@nus.edu.sg and yeesin\_ang@sutd.edu.sg}
\affiliation{Department of Physics, National University of Singapore, Singapore 117542}

\author{Yee Sin Ang}
\thanks{Authors to whom correspondence should be addressed: phylch@nus.edu.sg and yeesin\_ang@sutd.edu.sg}
\affiliation{Science, Mathematics and Technology (SMT), Singapore University of Technology and Design, Singapore 497372}

\begin{abstract}

Topological Dirac semimetals (TDSM), such as Cd$_3$As$_2$ and Na$_3$Bi, exhibits strong optical dichroism with contrasting dielectric permittivity along different crystal axes. However, such optical dichroism is often overlooked in the study of TDSM-based optoelectronic devices, and whether such optical dichroism can lead to unique functionalities not found under the isotropic approximation remain an open question thus far. Here we show that the optical dischroism in TDSM lead to starkly different terahertz (THz) responses and device performance as compared to the isotropic case. Using finite-difference time-domain simulations of a Cd$_3$As$_2$-based metasurface, we demonstrate that such optical dichroism can lead to an unexpected THz wave polarization conversion even if the metasurface structure remains C$_4$ four-fold rotationally symmetric, a practically useful feature not achievable under the isotropic model of TDSM. Our findings concretely reveal the contrasting spectral response between isotropic and anisotropic media, and shed important light on the capability of anisotropic TDSM in THz applications, leading not just to the more accurate device modelling, but also a new route in realizing THz waves polarization conversion without the need of complex device morphology commonly employed in conventional polarization converters.

%reveal a different perspective to achieve efficient and easy-to-implement electromagnetic polarizer by harnessing the optical anisotropy of TDSM. 
\end{abstract}
\maketitle

%\begin{introduction}

%\section{Introduction}

Three-dimensional (3D) topological Dirac semimetals (TDSM) represents peculiar 3D analogs of graphene whose electronic band structure around the Fermi level disperses linearly in all three orthogonal crystal directions.
Being a 3D bulk material with an additional spatial degree of freedom, TDSM is expected to offer greater device design flexibility not found in the atomically-thin graphene sheet while retaining the exotic physical properties of Dirac material systems. 
3D TDSMs, such as Cd$_3$As$_2$ and Na$_3$Bi, has been demonstrated to exhibit myriads of unusual transport and optical properties, such as giant magnetoresistance \cite{liang2015ultrahigh}, exceptionally high electrical mobility \cite{chanana2019manifestation,chorsi2020widely}, ultrafast relaxation dynamics \cite{lu2017ultrafast}, nonlinear plasmonics \cite{ooi2019nonlinear}, efficient high-harmonic generation \cite{zhang2019optical,ullah2020third,lim2021maximal,lim2020efficient,cheng2020efficient,kovalev2020non, cheng2020third,lee2020enhanced,tai2021anisotropic} %\textcolor{red}{[YS: doi.org/10.1038/s42005-021-00738-6, Phys. Rev. Res. 2, 043252 (2020), Phys. Rev. Lett. 124, 117402 (2020), Nat. Commun. 11, 2451 (2020).]}, 
high-temperature linear quantum magnetoresistance \cite{zhang2011topological,wang2021surface} and quantum Hall effect \cite{schumann2018observation,goyal2018thickness,zhang2017evolution}. 
Leveraging on these unusual physical properties, TDSM has been extensively explored as a promising material for a wide array of solid-state device applications \cite{wang2020topological} 
%\textcolor{red}{[YS: Cite doi.org/10.1021/acsnano.9b07990]},
including photodetector \cite{liu2020semimetals}, 
%\textcolor{red}{[YS: Cite doi.org/10.1038/s41563-020-0715-7]},
topological transistor \cite{collins2018electric},
%\textcolor{red}{[YS: Cite doi.org/10.1038/s41586-018-0788-5]},
spintronics \cite{suess2018topologically,wu2018dirac},
%\textcolor{red}{[YS: doi.org/10.1038/s41567-018-0064-5 and doi.org/10.1002/adma.201707547]}, 
metamaterials thermoelectric converters \cite{fu2020topological} 
%\textcolor{red}{[YS: doi.org/10.1063/5.0005481]} 
and electrical contacts to 2D semiconductor \cite{cao2020electrical}. %\textcolor{red}{[YS: Cite doi.org/10.1103/PhysRevApplied.13.054030]}.

TDSM is particularly attractive for nanophotonics applications \cite{dai2021high,chorsi2020widely} 
%\textcolor{red}{[YS: cite doi.org/10.1002/adfm.202011011 and doi.org/10.1002/adom.201901192]} 
in the terahertz (THz) frequency regime due to the gapless, and linear energy dispersion that are highly beneficial in generating strong THz optical nonlinearity. 
Computational studies of novel photonic and optoelectronic devices designs, such as optical switch \cite{zhu2017robust}, absorbers \cite{meng2019bidirectional,li2021multi}, photodetectors \cite{wang2017ultrafast,yang2018efficient,yang2018enhanced}, and polarization converters \cite{jia2021temperature,dai2019controllable}, have been intensively carried out in recent years. 
However, it is notwithstanding that most of these computational works are performed based on a simplistic isotropic Dirac semimetal conductivity model where the inherent \emph{Dirac band structure anisotropy} in realistic TDSM crystals, such as Cd$_3$As$_2$ and Na$_3$Bi, are neglected \cite{meng2019bidirectional,meng2018three,xiong2020dual} in which the carrier group velocity along two orthogonal directions can differ by one order of magnitude. 
The inclusion of Dirac band structure anisotropy appreciably modifies the optical properties of TDSM and dramatically transform TDSM into an anisotropic medium with strong optical dichroism \cite{lim2020broadband}. 
This aspect immediately leads to the following open questions that urgently need to be addressed:
What are the key differences between the simplistic isotropic model and the more realistic anisotropic model of TDSM in the computational design of photonic devices?  
Does the presence of optical dichroism in TDSM improve or degrade the performance of TDSM-based photonic devices?
More importantly, can optical dichroism in TDSM be employed to generate \emph{unique} functionalities not found in the isotropic counterpart?

%Given the profound phenomenological ramifications of optical isotropy in TDSM, it remains to be studied whether optical anisotropy can also drastically modify or improve the performances of TDSM-based optical devices, and thus, what are differences the optics response of anisotropic materials may bring comparing with that of isotropic materials and how much anisotropy can be harnessed for practical device applications remain an open question thus far. These questions could be useful for the design of photonic device. 
%Given the rapid developments of optoelectronic devices today, there is indeed a strong and urgent impetus for investigating the optical response of introducing the optical anisotropy into photonic system.
%Therefore, there is powerful reason to answer these questions so that researchers can understand more about the optical response of anisotropy material introduced into photonic system.   
%\cite{lin2017line}

%One of the most important applications associated with TDSM setups concerns the change of polarization state when the light through photonic system.

\begin{SCfigure*}
\includegraphics[width =1.5\linewidth]{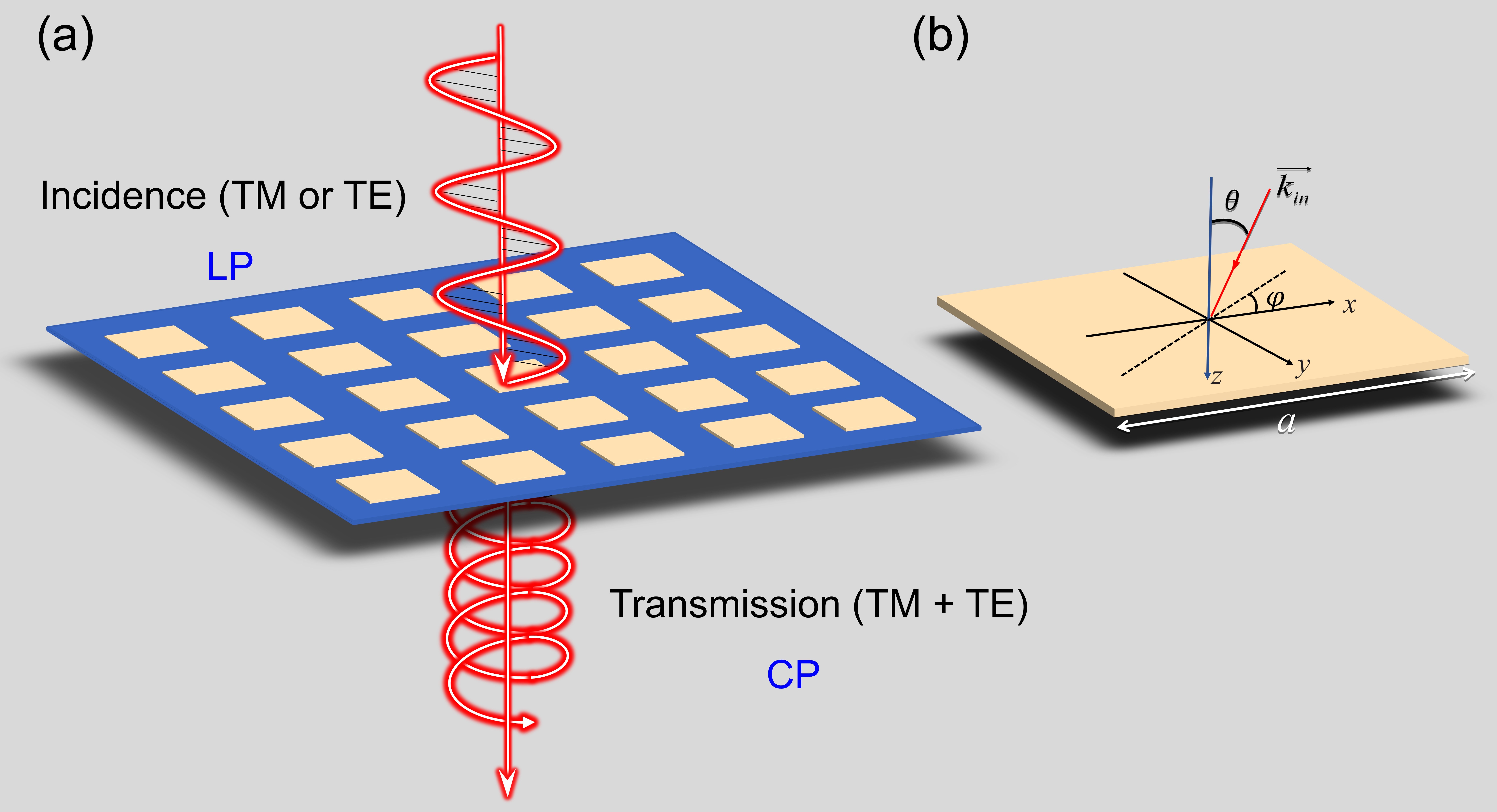}
\caption{\label{Fig:jhhpg} Device design of the Topological Dirac Semimetal Cd$_3$As$_2$ metasurface. (a) Schematic 3D view of the proposed Cd$_3$As$_2$ model system. The Cd$_3$As$_2$ layer has a thickness of 0.5 $\mu$m. %The electronic state of Cd$_3$As$_2$ is sensitive to its thickness: when the thickness is less than 50 nm, it is in the semiconductor phase; otherwise, it transforms into a Dirac semimetal phase. In our system, the Cd$_3$As$_2$ layer has thickness of 0.5 $\mu$m and it acts as a Dirac semimetal phase. 
%Plasmonic modes are supported in this system, which have amplitude and phase response characteristics. 
%Linearly-polarized wave illuminates the anisotropic Cd$_3$As$_2$ metasurface, resulting in asymmetric spectra response even in such a highly symmetric structure. With several special phase, the change of polarization state can be achieved. 
(b) One-pixel cell structure of the square-lattice-based metasurface. The pixels are arranged with periods $P_x=P_y = 86 \mu$m. The squares have a length of $a = 50$ $\mu$m. Polarization angle is defined as $\varphi$, with $\varphi= 0^\circ$ and $\varphi= 90^\circ$ correspond to TM and TE wave, respectively. A linearly-polarized plane wave impinges on the top of the system and propagates perpendicular to the $xy$-plane.}
\end{SCfigure*}

In this work, we address the above questions by performing a proof-of-concept device simulations of Cd$_3$As$_2$-based TDSM metasurface by explicitly taking into account the optical anisotropy of Cd$_3$As$_2$ crystal.   
We demonstrate that the far-field transmission spectra and the near-field electric field distributions (EFDs) are drastically different between the isotropic and the anisotropic TDSM model. Intriguingly, under the more realistic anisotropic TDSM model, a highly-symmetric C$_4$ patterns can readily generate THz wave polarization conversion - a peculiar behavior not achievable under the simplistic isotropic materials. These findings reveal a viable route to achieve efficient polarization conversion of THz waves with simple symmetric structures by harnessing the inherent optical dichroism of TDSM, thus circumventing the need of fabricating complex chiral sturctures \cite{hao2007manipulating,zhao2013tailoring,jun2017multi} or supercells composed of multiple sublayers and subunits \cite{yu2012broadband,yu2011light, kang2012wave, li2014ultra,han2018off, li2019multiple,gansel2009gold, shaltout2014optically}.

 We consider square arrays of Cd$_3$As$_2$ with a relative permittivity $\varepsilon$, thickness of 0.5 $\mu$m [see Fig. 1(a)]. 
Periodic boundary conditions are applied along both $x$ and $y$ directions with periodicity $P_x=P_y$ = 86 $\mu$m. The gap between adjacent square is 36 $\mu$m. 
%All of these parameters are fixed in the whole article. 
The coordinate system is defined such that x and y directions are lying in the plane of the square which has a width of $a = 50$ $\mu$m [Fig. 1(b)]. The finite-difference-time-domain (FDTD) method is employed to calculate transmission spectra, phase, amplitude and EFDs, which reveal the resonance mechanisms and transmission properties of the device. 
The polarization angle is defined as $\varphi$ as shown in Fig. 1(b). 
Thus, $\varphi = 0^\circ$ and $\varphi = 90^\circ$ correspond to TM- and TE-polarized wave, respectively. 
%For the system, $\varphi$ is meaningless to $\overrightarrow{k}_ {in}$, but is meaningful to differentiate the two polarizations (TE- and TM-polarized waves). 
%$\varphi$ is used to differentiate the two polarizations (i.e. TE-and TM-polarized waves). 
A linearly-polarized plane wave at normal incidence is considered. 
%The incidence plane contains the incidence ray and the normal  
%\textcolor{red}{[YS: What do you mean by 'normal ray'?]}. 
%\textcolor{blue}{[HY: I mean that the incidence ray and normal ray represent the red line and blue line in Fig.1(b), respectively.]}
To avoid nonphysical reflections of outgoing electromagnetic waves from the grid boundaries, a perfectly matched layer of absorbing boundary is employed along the propagation direction of the electromagnetic waves in the z direction. 
For computational simplicity, air is assumed to be the background. 
%Although it is a simplified model, it will be proved to be given useful hints on the optical response. 
 %( $\varepsilon_{xx}\ne\varepsilon_{yy}\ne\varepsilon_{zz}$)

%In recent years, the photoelectric devices based on TDSs have been intensively studied. However, most of TDSs devices were designed through the optical isotropy, while much less attention has been paid to optical anisotropy, which may make effect on the results when considering the optical response. 
The optical anisotropy of Cd$_3$As$_2$ is modelled based on a previously developed optical conductivity model obtained from Kubo formula \cite{lim2020broadband}. 
The semiclassical intraband and quantum mechanical interband optical conductivities are determined, respectively, as
\begin{subequations}
\begin{equation}
\sigma_{ii}^{\mathrm{intra}}(\omega) = \frac{ge^{2}v_{i}}{6\pi^{2}\hbar^{3}v_{j}v_{k}}\frac{\tau}{i\omega\tau-1}\int^{+\infty}_{-\infty}\frac{\partial f_{\mathrm{D}}(\mathcal{E})}{\partial\mathcal{E}}\mathcal{E}^{2}d\mathcal{E}
\label{eqn_intraband_sigma}
\end{equation}
\begin{equation}
\begin{split}
\sigma_{ii}^{\mathrm{inter}}(\omega) =& \frac{ige^{2}v_{i}\omega}{3\pi^{2}\hbar^{2}v_{j}v_{k}}\int^{+\infty}_{0}\frac{G(\mathcal{E})\mathcal{E}}{\hbar^{2}(\omega + i0^{+})^{2} - 4\mathcal{E}^{2}}d\mathcal{E},
\end{split}
\end{equation}
\end{subequations}
where $g = 4$ is the degeneracy factor, $i = x,y,z$ denotes the three orthogonal lattice directions, $v_i \neq v_j \neq v_k$ are the Fermi velocity parameter in the $i,j,k$-direction, respectively, $G(\mathcal{E}) = f_{\mathrm{D}}(-\mathcal{E}) - f_{\mathrm{D}}(\mathcal{E}) = \sinh(\beta\mathcal{E})/[\cosh(\beta\mathcal{E}) + \cosh(\beta\mu)]$, $f_{\mathrm{D}}(\mathcal{E})$ is the equilibrium Fermi-Dirac distribution function, $\mu = 50$ meV is the Fermi level, $\beta = 1/k_{\mathrm{B}}T$ is the inverse thermal energy, and $T = 300$ K at room temperature and the scattering time constant is assumed to be $\tau \approx 450$ fs. 
The permittivity can be calculated as $\varepsilon_{ii}(\omega) = 1 + \mathrm{i}\sigma_{ii}^{\mathrm{tot}}(\omega)/\omega_{\mathrm{p},i}\epsilon_{0}$, where $\epsilon_0$ is the permittivity of free space, and $\sigma_{ii}^{\mathrm{tot}} (\omega) \equiv \sigma_{ii}^{\mathrm{intra}}(\omega) + \sigma_{ii}^{\mathrm{inter}}(\omega)$.
For Cd$_3$As$_2$, the Fermi velocity parameter is highly anisotropic $(v_x, v_y, v_z) = (1.28, 1.30, 0.33) \times 10^6$ m/s. 
In this case, the permittivity exhibits a strong directional dependence with  $\varepsilon_{xx}\ne\varepsilon_{yy}\ne\varepsilon_{zz}$ in general. 
To highlight the significance of optical anisotropy of TDSM, we also perform benchmark studies using the isotropic optical model of TDSM \cite{liu2018dirac,chen2017realization,cai2020tunable,wang2020tunable}. 
%\textcolor{red}{[YS: add some references that uses isotropic model to simulate TDSM optical devices as some example here]}
As demonstrated below, the C$_4$-symmetric metasurface studied here exhibit radically different optical responses when the optical anisotropy is omitted, thus highlighting the importance and necessity of incorporating the anisotropic permittivity in the computational designs of TDSM-based photonic devices. 

%To motivate our study and emphasize the effects of permittivity anisotropy in Cd$_3$As$_2$, we also present results on the hypothetical scenario where Cd$_3$As$_2$ has isotropic permittivity, with exactly the same setup. Notably, the electronic state of Cd$_3$As$_2$ is sensitive to its thickness: when the thickness is less than 50 nm, the semiconductor phase dominates; otherwise, it transforms into a Dirac semimetal phase. In our design system and THz region, the thickness of Cd$_3$As$_2$ is set as 0.5 $\mu$m, which is a Dirac semimetal phase. 
%TDSs have been manifested to support plasmons in the THz frequency range, thus it enables the study of plasmonic excitations in our system. 

%\section{Results and Discussions}
%\subsection{Comparison of transmission spectra between isotropic and anisotropic case}

\begin{figure*}
\includegraphics[width=1\linewidth]{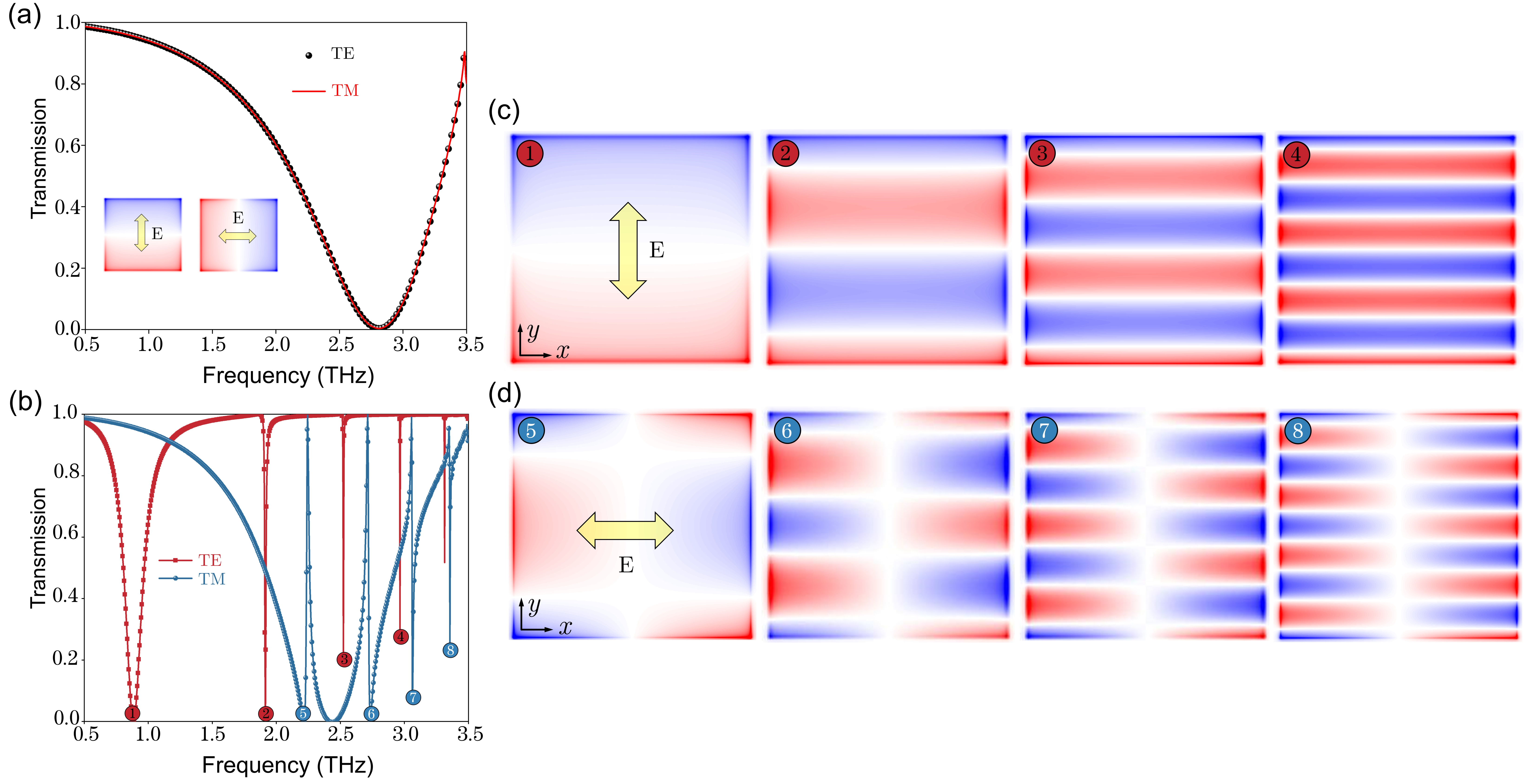} %[scale=0.2058]
\caption{\label{Fig:jpg} 
%\textcolor{red}{[YS: This figure has a strip of whitespace below the figure, please remove this white space, so to conserve the page length.]} 
Comparison of the transmission spectra and electiric field distributions between the isotropic and anisotorpic optical models. (a) Simulated transmission spectra for the isotropic case under TE-and TM-polarized wave. The transmission signal shows practically identical for both TE and TM polarizations. Inset shows the electric field distributions (EFDs) for the two polarizations. (b) Transmission spectra for the anisotropic case under TE-and TM-polarized wave. We labeled the first four modes as \ding{172}-\ding{175} and \ding{176}-\ding{179} for TE-and TM-polarized wave, respectively. 
%The resonance frequencies and intensities are all different from each other, which indicates a strong dichroism. 
(c) and (d) shows the surface EFDs on the top surface of the metasurface structure for symmetric and antisymmetric modes, respectively.}
\end{figure*}

In symmetric structure with isotropic materials, super-radiant or bright modes couple strongly to the incident field, producing broad and lossy resonances. In the presence of geometrical asymmetry, sub-radiant or dark modes can be excited. These dark mode resonances couple weakly to the free space and lead to high values of quality factor ($Q$-factor). Fano-type resonance caused by symmetry breaking can result in an asymmetric spectral profile. In relevance to this, we first focus on the optical response of Cd$_3$As$_2$ metasurface outlined in Fig. 1(a) under the \emph{oversimplified} isotropic approximation of $\varepsilon_{xx}=\varepsilon_{yy}=\varepsilon_{zz}$. 
In this case, the transmission response is identical under both TE- and TM-polarized wave [Fig. 2(a) and the inset for the corresponding EFDs]. 
In contrast, when the optical anisotropy is included, the transmission spectra exhibit completely different line shapes for TE- and TM-polarized waves [Fig. 2(b)]. 
For a better insight into the nature of these resonant modes, we plot the out-of-plane EFDs (i.e. $E_z$) at each resonance mode [Figs. 2(c) and 2(d)].
The EFDs of the first four modes, labelled as \ding{172}-\ding{175} in Fig. 2(b), are shown in Fig. 2(c) for TE-polarized wave. 
A broadband feature occurs at about 0.88 THz [labeled as \ding{172}] corresponds to the super-radiant mode that couples strongly to the free space. The radiation field of the dipole interferes strongly, resulting in the occurrence of various symmetric dipole modes [see Fig. 2(c)]. 
In addition to these dipolar modes, four narrower asymmetric resonance modes, labeled as \ding{176}-\ding{179} in Fig. 2(b), which correspond to resonances with very low radiation losses and high \emph{Q} factor, are also present under the TM-polarized wave. 
Furthermore, the symmetric modes in Fig. 2(c) display a net dipole moment. 
Such net dipole moment is completely absent in the asymmetric modes in Fig. 2(d). 
Importantly, Fig. 2 reveals the very different nature of the resonances under TE- and TM-polarized incidence. 
Such phenomenon is a direct consequence of the anisotropic permittivity of optically anisotropic Cd$_3$As$_2$. Thus, by exciting these different spectral responses, one can switch between the symmetric and antisymmetric effects simply through rotating the polarization direction of the incident light.

%It can be clearly seen from Fig. 2(b) that in striking contrast with the TE-polarized wave, the transmission spectra under the TM-polarized shows a pronounced difference in transmitted intensity and resonance position. This is distinct from the isotropic case, where the spectra show the same characteristic under the two orthogonal polarizations. 
%As elaborated below, the transmission signal still shows several resonance modes and these modes also possess narrow broadband and couple rather inefficiently to external light, labeling as \ding{176}-\ding{179}. As the EFDs shown in Fig. 2(d), these modes are antisymmetric modes. Up to now, we can indicate that when a plane wave is placed above the square, symmetric modes can be excited with polarization parallel y axis, whereas antisymmetric modes couple to the TM-polarized wave. This difference in transmittance for the two polarization states indicates the presence of a strong dichroism. 

%, which is much different from the polarization-insensitive spectra effect (see discussion in Fig. 2(a)). 
%Although we use square structure in this simulation, our results can be easily extended to other structures with highly symmetry such as disk and do not lose generality. And differences just the resonance frequency and reflection intensity, but the electric fields will keep the same distribution as this case. 

\begin{table*}
\caption{Comparison of isotropic and anisotropic modelling of Cd$_3$As$_2$ square-lattice metasurface. The resonant frequency, FWHM, $Q$-factor and the corresponding mode label displayed in Fig. 2 are shown.}
\begin{ruledtabular}
\begin{tabular}{ccccc}
Systems & Resonant frequency (THz) &
\multicolumn{1}{c}{\textrm{Mode label}} &
\multicolumn{1}{c}{\textrm{FWHM (THz)}} &
\multicolumn{1}{c}{\textrm{$Q$-factor}} \\
%\mbox{Three}&\mbox{Four}&\mbox{Five}\\
\hline
isotropic TDSM (TE)&2.76&\mbox{n.a.}&\mbox{1.17}&\mbox{2.36}\\
\hline
isotropic TDSM (TM)&2.76&\mbox{n.a.}&\mbox{1.17}&\mbox{2.36}\\
\hline
%\multirow{4}{*}{isotropic Cd$_3$As$_2$ (TE)}
anisotropic TDSM (TE)& 0.88 & \ding{172} & 0.17 & 5.17\\
& 1.91 & \ding{173} & 0.01 & 191\\
& 2.53 & \ding{174} & 0.04 & 632.5\\
& 2.96 & \ding{175} & 0.03 & 986.7\\
\hline
anisotropic TDSM (TM) & 2.21 &\ding{176} & 0.03 & 73.6\\
& 2.73 & \ding{177} & 0.019 & 143.7\\
& 3.06 & \ding{178} & 0.08 & 382.5\\
& 3.36 & \ding{179} & 0.07 & 480\\
\end{tabular}
\end{ruledtabular}
\end{table*}

\begin{figure*}
\includegraphics[scale=0.4758]{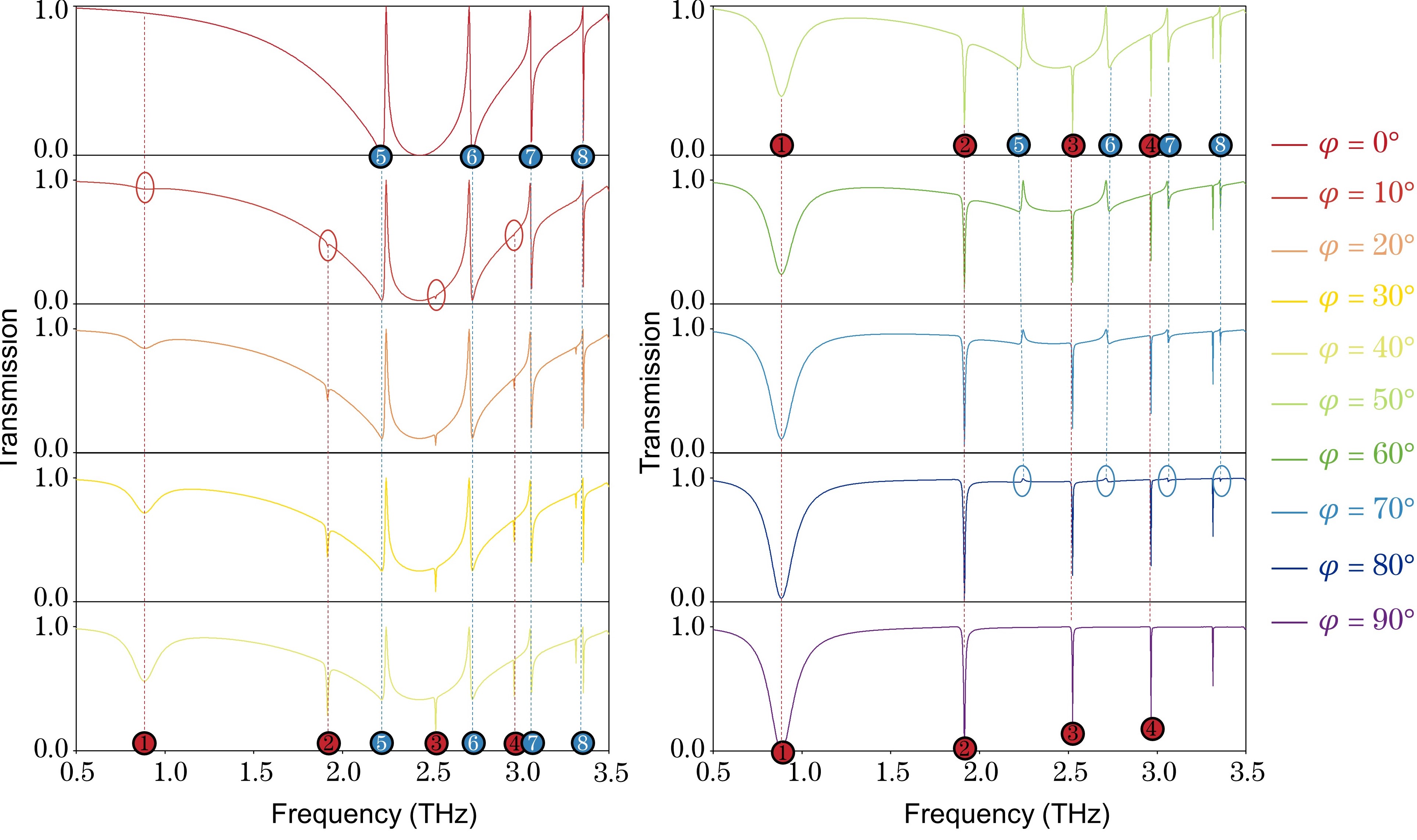}
\caption{\label{Fig:jpgtttt} Transmission spectra for different polarization angles of the incident wave. 
%This transmission feature show that by changing the polarization state of linearly-polarized normally incident on the plane, the strength of the transmitted modes and intensities will change.
The modes labelled \ding{172} to \ding{179} here are the same as that labelled in Fig. 2. As $\varphi$ increases from $0^\circ$ towards $90^\circ$, the sharp transmission peaks corresponding to the antisymmetric modes (denoted by the labels \ding{172} to \ding{175} in Fig. 3) recede gradually while that corresponding to symmetric modes [denoted by the labels \ding{176} to \ding{179} in Fig. 3] become increasingly prominent, showing a direct consequence of the optical dichroism of Cd$_3$As$_2$.}
\end{figure*}

The device performance parameters, namely the resonant frequency, full width at half maximum (FWHM) and $Q$-factor, are listed in Table I for the metasurface simulated under both isotropic and anisotropic models.  Interestingly, the FWHM and $Q$-factor of isotropic approximation is significantly smaller than that of the anisotropic optical model. The exceptionally high $Q$-factor, reaching as high as 10$^2$ to $10^3$, obtained from the anistropic model suggests an excellent performance in THz region failed to be captured under the isotropic approximation. 
The underlying reason of the better performance is due to the significantly Fermi velocity of $1.30\times 10^6$ m/s along the y crystal direction of Cd$_3$As$_2$ %\textcolor{red}{[YS: Haiyu, please check Jeremy's paper to find the value of the 'large' Fermi velocity and the corresponding direction]} 
is significantly larger than that commonly employed in the isotropic approximation of $1.0\times 10^6$ m/s %\textcolor{red}{[YS: Please also quote the Fermi velocity commonly used in the isotropic TDSM model]}, 
which directly influences the $Q$-factor. 
In general, larger Fermi velocity constant produce the resonances with a larger $Q$-factor.
This can be seen from the optical conductivity expression in Eq. (1).
The $i$-directional conductivity is generally inversely proportional to $v_{j,k}$. 
A larger Fermi velocity constant thus suppresses the real part of the optical conductivity and, correspondingly, the imaginary part of the dielectric permittivity, $\Im[ \varepsilon(\omega)]$. 
The suppression of $\Im[ \varepsilon(\omega)]$ results in the exceptionally large $Q$-factor. 
%Let us now turn to the comparison of inside anisotropic Cd$_3$As$_2$. 
The narrow FWHM and high $Q$-factor typically occurs for higher modes \ding{173}-\ding{175}, %\textcolor{red}{[YS: Explicitly quote the mode label numbers here]},
suggesting that these sharp peaks and dips of Fano resonance may be harnessed for high-sensitivity biosensors applications. 
%Therefore, it is expected that a variety of sharp resonance spectra for applications of sensing and filters. 
Importantly, the device performance characteristics outlined in Table I suggest that the anisotropic optical model of TDSM better captures the superior application potential  of Cd$_3$As$_2$ in THz waves modulation, as compared to the less accurate, or oversimplified, isotropic optical model.

\begin{figure*}[t]
\includegraphics[width=1\linewidth]{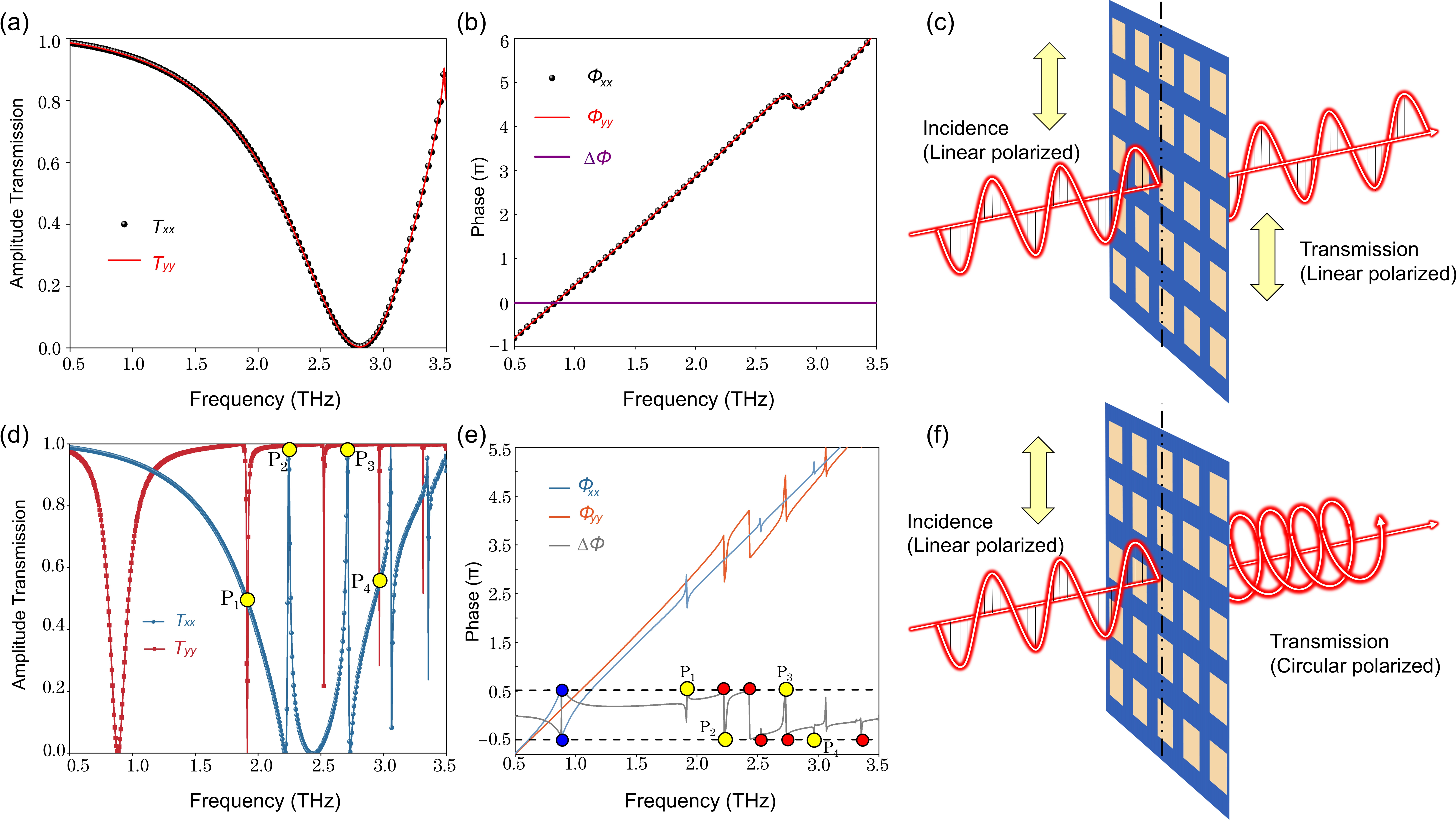}
\label{Fig:jpg1}
\caption{Transmission amplitude and relative phase delay of the isotropic and anisotropic TDSM metasurface. (a) Transmission amplitudes, $T_{xx}$ and $T_{yy}$, of the isotropic models of TDSM metasurface as a function of frequency, where $T_{xx}$ and $T_{yy}$ are the x-polarized transmission under x-polarized incidence, and y-polarized transmission under y-polarized incidence, respectively. (b) Phase information, $\phi_{xx}$, $\phi_{yy}$, and the relative phase delay $\Delta\phi$ of the isotropic models of TDSM as a function of frequency, where $\phi_{xx}$ and $\phi_{yy}$ are the phases of the transmitted waves along the x- and y-directions, respectively. (c) Schematic illustration of THz wave transmission across an isotorpic TDSM metasurface in which polarization conversion effect is absent. (d) and (e), same as (a) and (b), but for the case of optically anisotropic TDSM. 
%It is observed that the phase difference response shows nonmonotonic behavior with frequency, and it can take values within [-$\pi$/2, $\pi$/2] depending on the frequency. Under this condition, when $T_{xx}$ and $T_{yy}$ are in a special relationship, the possible polarization states (circular, linear, elliptic) are realizable. The four green points and seven red points represent the relative phase delay is -$\pi$/2 or $\pi$/2, of which the four green points refer to position $T_{xx}$ = $T_{yy}$ and the seven red points refer to $T_{xx} \ne T_{yy}$. 
(f) Schematic illustration of the linear-to-circular polarization conversion in an optically dichroic TDSM metasurface.}
\end{figure*}

%\subsection{Transmission spectra with different polarization angles. }

To gain an understanding on how different polarization angles can influence the transmission intensity and spectra position, the optical response at varying $\varphi$ is calculated. 
We calculate the transmission spectra of the system at several polarization angles ranging from $\varphi$ = $0^\circ$ to $\varphi$ = $90^\circ$ with $10^\circ$ steps. Two different processes are responsible for the observed variation in transmission spectra shown in Fig. 3. 
The incident light with an arbitrary polarization angle $\varphi$ can be decomposed into two components, namely the parallel ($\varphi$ = $0^\circ$) and perpendicular ($\varphi$ = $90^\circ$) polarization components. 
%The transmission spectra for incident light with $\varphi$ = $0^\circ$ and $\varphi$ = $90^\circ$ correspond to the excitation of antisymmetric and symmetric modes, respectively. 
For an incident wave with $\varphi$ = $0^\circ$ and $\varphi = 90^\circ$, only the antisymmetric  and the symmetric modes are excited, respectively. 
However, when $\varphi$ is intermediate between $0^\circ$ and $90^\circ$, the spectra response exhibit a mixture of the symmetric and antisymmetric modes. 
As $\varphi$ increases from $0^\circ$ towards $90^\circ$, the sharp transmission peaks corresponding to the antisymmetric modes (denoted by the labels \ding{172} to \ding{175} in Fig. 3) recede gradually while that corresponding to symmetric modes [denoted by the labels \ding{176} to \ding{179} in Fig. 3] become increasingly prominent. 
%As the polarization angles further increase, symmetric modes are gradually visible, while the antisymmetric modes gradually decrease. 
%This phenomenon can be attributed to the fact that the component along y-direction of electric field gradually increase as $\varphi$ increases. 
%Conversely, when y-polarized wave ($\varphi$ = $90^\circ$) was applied, the direction of electric field of the incidence light totally along y axis and no any elements along x axis, leading to the appearance of only symmetric modes and antisymmetric modes totally disappear. 

The $\varphi$-tunable symmetric-to-antisymmetric mode transition reveals an intriguing behavior of optically anisotropic TDSM metasurface. The transmitted electromagnetic waves can be dramatically modulated by changing the polarization angle of a linearly polarized incident wave even though the metausrface is composed of highly symmetric square lattices.
This behavior is a direct consequence of the optical dichoism of Cd$_3$As$_2$ and is completely missed out in the simplistic isotropic optical conductivity model of TDSM. 

%Benefiting from the natural optical anisotropy of Cd$_3$As$_2$, the transmission coefficient of light varies with the directions of polarization when the light propagates in the system. 
%For this optical anisotropic system, this asymmetric effect can be explained by the following discussion. 

The occurrence of highly asymmetric spectra response from a highly symmetric structure reveals the potential of Cd$_3$As$_2$ -- an optically anisotropic TDSM -- in achieving efficient polarization conversion without involving complex chiral or supercell structures that are challenging to fabricate experimentally. 
At normal incidence, the incident electric field can be represented by $\overrightarrow{E}^{in}=(E_x\hat{x}+E_y\hat{y})e^{i(\omega z/c)}$, where $E_x$ and $E_y$ are the electric field components along the $x$ and $y$ directions, respectively.
After traversing a high-symmetry structure and isotropic system, the transmitted electric field becomes $\overrightarrow{E}^{t}_{iso}=t(E_x\hat{x}+E_y\hat{y})e^{i(\omega z/c+\omega t)}$, where $t$ is the transmission coefficient, and the transmitted electric field is insensitive to the relative contributions between $E_x$ and $E_y$ for a given incident wave. 
In contrast, for an anisotropic system with $\varepsilon_{xx}\ne\varepsilon_{yy}\ne\varepsilon_{zz}$, the transmission coefficients are different for incident waves polarized along the $x$ and $y$ directions. The transmitted electric field becomes $\overrightarrow{E}^{t}_{aniso}=(t_xE_x\hat{x}+t_yE_y\hat{y})e^{i(\omega z/c+\omega t)}$ with \emph{directional-dependent} transmission coefficients, $t_x$ and $t_y$. As $t_x \neq t_y$ in general, the relative composition of $E_x$ and $E_y$ of the incident wave can dramatically modulates the transmitted wave. As demonstrated below, the optical anisotropy of TDSM can be readily harnessed to achieve linear-to-circular polarization conversion of an incident THz wave -- a behavior not achievable under the isotropic model of TDSM.

%\subsection{Realization of changing polarization state using the dichroism} 

%It is intriguing that the optical response can behave distinctly under orthogonal polarization, and thus, may realize desirable application. 
%In previous works, the asymmetric effect can be realized generally by designing a metamaterial composed of chiral meta-atoms. 
%In our work, asymmetric spectra effect is realized in a C$_4$ rotational symmetry structure. Such a property is distinctly different from the previous chiral metasurface designs to achieve an asymmetric reflection or transmission. 

The phase and amplitude play a crucial role in determining the polarization state of the transmitted wave. 
%To understand this significance of strong dichroism behavior of the C$_4$ structure, 
To characterize the polarization conversion operation of TDSM, we calculate the phase and amplitude of the transmission spectra in Fig. 4.
Here, polarization conversion is related to the relative phase delay between the $x$- and $y$-components of the transmitted wave. 
%Based on this relationship, we will show how our metasurface achieve the conversion using this dichroism. 
%In the previous parts, we have made clear the optics response of isotropic and anisotropic materials structured by C$_4$ symmetric square. 
We first show that, under the isotropic approximation, TDSM is not capable of polarization conversion.
In this case, the transmission amplitudes, $T_{xx}$ and $T_{yy}$ [Fig. 4(a)] and the phases, $\phi_{xx}$ and $\phi_{yy}$, of the $x$- and $y$-components of the transmitted wave are completely identical [Fig. 4(b)]. 
Thus, the transmitted wave retains the polarization state of the incident wave in an isotropic TDSM model[Fig. 4(c)], which is expected due to the highly-symmetric nature of the square lattices exhibiting a C$_4$ rotational symmetry perpendicular to the optical axis of the normal incidence. 
%In addition to the expected difference spectra property, the main difference between anisotropic and the isotropic case concerns the phase difference, as it will be discussed thoroughly later. 
%We first present the simulated amplitude transmission and phase spectra with isotropic case [see Figs. 4(a) and (b)], and then give the schematic of the changes to the polarization state of light transmitted through the metasurface [see Fig. 4(c)]. 
%As shown in Fig. 4(a), $T_{xx}$ and $T_{yy}$ are the x-polarized transmission under x-polarized incidence, y-polarized transmission under y-polarized incidence, respectively. 
%As can be seen, $T_{xx}$ and $T_{yy}$ show the same curve and the relative phase delay is zero, which means that the C$_4$ symmetric structure under isotropic case displays no discernible polarization state conversion, as the schematic shown in Fig. 4(c). 
%This behavior is expected as the geometry exhibits a C$_4$ rotational symmetry perpendicular to the optical axis for light at normal incidence. 
In contrast, transmission amplitudes are drastically different for TE- and TM-polarized waves in the case of anisotropic TDSM [Fig. 4(d)]. 
The corresponding $\phi_{xx}$ and $\phi_{yy}$ exhibit contrasting dispersion between the $x$- and $y$-components of the transmitted waves [Fig. 4(e)]. 
Aside from amplitude parameter, the phase difference defined as $\Delta\phi=\phi_{xx}-\phi_{yy}$ is another crucial role for the polarization conversion. As shown in Fig. 4(e), the phase difference distinctly indicate a strongly nonlinear curve. Specifically, the mode \ding{172} located at 0.88 THz lifts the phase dispersion curve and a phase shift of $\pi$ occurs. 
Intriguingly, the right-side tail of mode \ding{172} (0.88 THz) and the left-side tail of mode \ding{173} (1.91 THz) makes a close-to-linear dispersion at off-resonance frequencies between the two modes. Simultaneously, the same tendency appears between other adjacent resonance modes with different phase jump. Such a nonlinear phase dispersion related to the polarization state of transmission wave provides a possibility of efficient-broadband polarization conversion. 
Importantly, the relative phase delay is limited to the range of $-\pi/2$ and $+\pi/2$, covers multiple special values, such as -$\pi$/2, 0 and $\pi$/2, thus indicating that multiple polarization states (i.e. circularly-, linearly-, elliptically-polarized wave) can be achieved by appropriately matching the transmitted wave amplitudes and the relative phase delay.   
For example, linear-to-circular polarization conversion, akin to the polarization conversion of a quarter-wave plate, can be achieved with the simultaneous fulfilment of $T_{xx}$ = $T_{yy}$ and $\Delta\phi$ = $\pm \pi$/2 which generates two co-propagating wave components of equal amplitudes of orthogonal linear polarizations and phase-shifted by a quarter-wavelength [Fig. 4(f)]. 
Within the frequency window of 0.5 THz to 3.5 THz, there are four operation frequencies [labelled as $P_1$ to $P_4$ in Figs. 4(d) and 4(e)] capable of producing linear-to-circular polarization conversion. 
%The first resonance occurs at 0.88 THz and 2.23 THz for y-polarized wave and x-polarized wave, respectively. Between the two resonances, the phase spectra are nearly linear the region from 0.88 to 1.91 THz, as shown in Fig. 4(e). This provides a nearly constant phase difference 0.2$\pi$, but unfortunately, the phase difference does not facilitate a circular-polarized wave. 
%It can be found that $\Delta\phi=-\pi/2$ and $\pi$/2 at green and red points in Figs. 4(e). 
%As a result of these two properties, upon the excitation by linearly-polarized light, generating two co-propagating waves with equal amplitudes, orthogonal linear polarizations, which means that the transmission waves produce a circularly-polarized state. 

Apart from linear-to-circular polarization conversion, \emph{linear-to-elliptical} polarization conversion can also be achieved under the conditions of $\Delta\phi = n\pi/2$ but with $T_{xx} \ne T_{yy}$. 
Multiple of such operating frequencies are identified as labelled by the red circles in Figs. 4(d) and 4(e). 
Furthermore, between the two resonances frequencies of 0.88 THz to 1.91 THz, an expansive region of nearly constant phase delay of $\sim 0.2\pi$ is present, suggesting a broadband frequency windows capable of producing linear-to-elliptical polarization conversion. 
%\textcolor{red}{[YS: There is a near-constant 0.2$\pi$ phase shift region between 0.8 THz and 1.91 THz - is there any significance of this constant phase shift region? What kind of useful polarization conversion can they produce?]
%\textcolor{blue}{[HY: 0.2$\pi$ will produce elliptic-polarized wave]}
%\textcolor{red}{[YS: How about the case of $\pm \pi/2$ phase shift, but $T_{xx} \neq T_{yy}$? Does this also produce elliptical polarization?]}
Importantly, despite the relatively simple geometry of the square-lattice-based metasurface, the optical dichroism of Cd$_2$As$_3$ enables such metasurface to perform various polarization conversions in THz regime. This highlights the importance and unique features of optically dichroic TDSM not captured by the isotropic optical model commonly employed in the literature.
Finally, we note that the geometrical parameters of the Cd$_3$As$_2$ metasurface can be further optimized the achieve desired polarization conversion performance. 

%It will be straightforward just by simply rotating the metasurface or the incidence linear polarization by 90°. 
%Comparing with the designed structures with successive layers in a third dimension to rotate the angular offset \cite{gansel2009gold}, our design is single-layered, which is always required for the development of integrated and miniaturized optical systems and simple than multilayer structure design. In addition, the C$_4$ symmetric structure simplifies the overall structure when comparing to designs with chiral metasurface structures \cite{zhao2013tailoring,jun2017multi} and the structural supercells with location and orientation of individual antennas \cite{yu2012broadband,shaltout2014optically}, indicating the feasibility and advantages of anisotropic Cd$_3$As$_2$-based THz polarizers. 

%\section{Conclusions}
In conclusion, we reported the optical response and polarization conversion operation of a C$_4$ symmetric metasurface composed of Cd$_3$As$_2$ square lattices. 
By explicitly taking into account the optical dichroism of Cd$_3$As$_2$ using an optical conductivity model obtained from the Kubo formula \cite{lim2020broadband}, we found that the metasurface exhibits strongly asymmetric transmission spectra sensitively influenced by the polarization angle of the incident waves. 
Such directional-dependent transmission dramatically affects the optical response of the metausrface and is not captured by the isotropic TDSM model. 
Intriguingly, the optical dichroism of Cd$_3$As$_2$ can be harnessed to achieve linear-to-circular and linear-to-elliptical polarization conversion of terahertz waves using a simple square-lattice metasurface, thus circumventing the need of complex device morphology, such as chiral or multilayer structures. 
These findings revealed the shortcomings of the commonly employed isotropic optical model of TDSM in the computational modeling of TDSM-based photonic devices, and highlighted the unique strength of optically anisotropic topological Dirac semimetal in THz waves polarization conversion applications without the need of complex device morphology.

%metasurface and the designed system exhibits distinct asymmetric transmission subjected to the orthogonal polarization states in contrast to the hypothetical isotropic Cd$_3$As$_2$, indicating a strong dichroism. We demonstrated that this single sheet can efficiently control the phase while maintaining a high transmission. In particular, a linearly-polarized light could converted into circularly-, elliptically- and linearly-polarized light after transmission by the anisotropic Cd$_3$As$_2$ metasurface under certain conditions, offering a new gateway to achieving advanced THz polarizers. Our study draw the optical response of anisotropic material and highlights optical anisotropic Cd$_3$As$_2$ as prospective constituents for THz polarizer applications. 

\begin{acknowledgements}
Y.S.A. is supported by the Singapore University of Technology and Design (SUTD) Start-up Research Grant (SRG SCI 2021 163). 
%C.H.L. is supported by WBS:R-144-000-435-133. 
H.Y.M. is supported by the China Scholarship Council (Grant No.202006130068)and the National Natural Science Foundation of China (61775055).%XXXXXXXXXXXX and YYYYYYYYYYYY \textcolor{red}{[Haiyu, please include your funding, such as CSC and your Hunan university scholarship]}
\end{acknowledgements}

\section*{Author Declarations}

\subsection*{Conflict of Interest}
\noindent The authors declare that there are no conflicts of interest.

%\subsection*{Author Contributions}
%\noindent H. Y. M. performed the calculation and analysis. H. Y. M. and Y. S. A. conceptualized and initiated the project. Y. S. A. and C. H. L. supervised the project. All authors contributed to the writing of this work.

\section*{Data Availability}
The data that support the findings of this study are available from the corresponding author upon reasonable request.

\bibliographystyle{apsrev4-2}
\bibliography{Reference}

%apsrev4-2.bst 2019-01-14 (MD) hand-edited version of apsrev4-1.bst
%Control: key (0)
%Control: author (72) initials jnrlst
%Control: editor formatted (1) identically to author
%Control: production of article title (-1) disabled
%Control: page (0) single
%Control: year (1) truncated
%Control: production of eprint (0) enabled
\begin{thebibliography}{53}%
\makeatletter
\providecommand \@ifxundefined [1]{%
 \@ifx{#1\undefined}
}%
\providecommand \@ifnum [1]{%
 \ifnum #1\expandafter \@firstoftwo
 \else \expandafter \@secondoftwo
 \fi
}%
\providecommand \@ifx [1]{%
 \ifx #1\expandafter \@firstoftwo
 \else \expandafter \@secondoftwo
 \fi
}%
\providecommand \natexlab [1]{#1}%
\providecommand \enquote  [1]{``#1''}%
\providecommand \bibnamefont  [1]{#1}%
\providecommand \bibfnamefont [1]{#1}%
\providecommand \citenamefont [1]{#1}%
\providecommand \href@noop [0]{\@secondoftwo}%
\providecommand \href [0]{\begingroup \@sanitize@url \@href}%
\providecommand \@href[1]{\@@startlink{#1}\@@href}%
\providecommand \@@href[1]{\endgroup#1\@@endlink}%
\providecommand \@sanitize@url [0]{\catcode `\\12\catcode `\$12\catcode
  `\&12\catcode `\#12\catcode `\^12\catcode `\_12\catcode `\%12\relax}%
\providecommand \@@startlink[1]{}%
\providecommand \@@endlink[0]{}%
\providecommand \url  [0]{\begingroup\@sanitize@url \@url }%
\providecommand \@url [1]{\endgroup\@href {#1}{\urlprefix }}%
\providecommand \urlprefix  [0]{URL }%
\providecommand \Eprint [0]{\href }%
\providecommand \doibase [0]{https://doi.org/}%
\providecommand \selectlanguage [0]{\@gobble}%
\providecommand \bibinfo  [0]{\@secondoftwo}%
\providecommand \bibfield  [0]{\@secondoftwo}%
\providecommand \translation [1]{[#1]}%
\providecommand \BibitemOpen [0]{}%
\providecommand \bibitemStop [0]{}%
\providecommand \bibitemNoStop [0]{.\EOS\space}%
\providecommand \EOS [0]{\spacefactor3000\relax}%
\providecommand \BibitemShut  [1]{\csname bibitem#1\endcsname}%
\let\auto@bib@innerbib\@empty
%</preamble>
\bibitem [{\citenamefont {Liang}\ \emph {et~al.}(2015)\citenamefont {Liang},
  \citenamefont {Gibson}, \citenamefont {Ali}, \citenamefont {Liu},
  \citenamefont {Cava},\ and\ \citenamefont {Ong}}]{liang2015ultrahigh}%
  \BibitemOpen
  \bibfield  {author} {\bibinfo {author} {\bibfnamefont {T.}~\bibnamefont
  {Liang}}, \bibinfo {author} {\bibfnamefont {Q.}~\bibnamefont {Gibson}},
  \bibinfo {author} {\bibfnamefont {M.~N.}\ \bibnamefont {Ali}}, \bibinfo
  {author} {\bibfnamefont {M.}~\bibnamefont {Liu}}, \bibinfo {author}
  {\bibfnamefont {R.}~\bibnamefont {Cava}},\ and\ \bibinfo {author}
  {\bibfnamefont {N.}~\bibnamefont {Ong}},\ }\href@noop {} {\bibfield
  {journal} {\bibinfo  {journal} {Nature materials}\ }\textbf {\bibinfo
  {volume} {14}},\ \bibinfo {pages} {280} (\bibinfo {year} {2015})}\BibitemShut
  {NoStop}%
\bibitem [{\citenamefont {Chanana}\ \emph {et~al.}(2019)\citenamefont
  {Chanana}, \citenamefont {Lotfizadeh}, \citenamefont {Condori~Quispe},
  \citenamefont {Gopalan}, \citenamefont {Winger}, \citenamefont {Blair},
  \citenamefont {Nahata}, \citenamefont {Deshpande}, \citenamefont
  {Scarpulla},\ and\ \citenamefont
  {Sensale-Rodriguez}}]{chanana2019manifestation}%
  \BibitemOpen
  \bibfield  {author} {\bibinfo {author} {\bibfnamefont {A.}~\bibnamefont
  {Chanana}}, \bibinfo {author} {\bibfnamefont {N.}~\bibnamefont {Lotfizadeh}},
  \bibinfo {author} {\bibfnamefont {H.~O.}\ \bibnamefont {Condori~Quispe}},
  \bibinfo {author} {\bibfnamefont {P.}~\bibnamefont {Gopalan}}, \bibinfo
  {author} {\bibfnamefont {J.~R.}\ \bibnamefont {Winger}}, \bibinfo {author}
  {\bibfnamefont {S.}~\bibnamefont {Blair}}, \bibinfo {author} {\bibfnamefont
  {A.}~\bibnamefont {Nahata}}, \bibinfo {author} {\bibfnamefont {V.~V.}\
  \bibnamefont {Deshpande}}, \bibinfo {author} {\bibfnamefont {M.~A.}\
  \bibnamefont {Scarpulla}},\ and\ \bibinfo {author} {\bibfnamefont
  {B.}~\bibnamefont {Sensale-Rodriguez}},\ }\href@noop {} {\bibfield  {journal}
  {\bibinfo  {journal} {ACS nano}\ }\textbf {\bibinfo {volume} {13}},\ \bibinfo
  {pages} {4091} (\bibinfo {year} {2019})}\BibitemShut {NoStop}%
\bibitem [{\citenamefont {Chorsi}\ \emph {et~al.}(2020)\citenamefont {Chorsi},
  \citenamefont {Yue}, \citenamefont {Iyer}, \citenamefont {Goyal},
  \citenamefont {Schumann}, \citenamefont {Stemmer}, \citenamefont {Liao},\
  and\ \citenamefont {Schuller}}]{chorsi2020widely}%
  \BibitemOpen
  \bibfield  {author} {\bibinfo {author} {\bibfnamefont {H.~T.}\ \bibnamefont
  {Chorsi}}, \bibinfo {author} {\bibfnamefont {S.}~\bibnamefont {Yue}},
  \bibinfo {author} {\bibfnamefont {P.~P.}\ \bibnamefont {Iyer}}, \bibinfo
  {author} {\bibfnamefont {M.}~\bibnamefont {Goyal}}, \bibinfo {author}
  {\bibfnamefont {T.}~\bibnamefont {Schumann}}, \bibinfo {author}
  {\bibfnamefont {S.}~\bibnamefont {Stemmer}}, \bibinfo {author} {\bibfnamefont
  {B.}~\bibnamefont {Liao}},\ and\ \bibinfo {author} {\bibfnamefont {J.~A.}\
  \bibnamefont {Schuller}},\ }\href@noop {} {\bibfield  {journal} {\bibinfo
  {journal} {Advanced Optical Materials}\ }\textbf {\bibinfo {volume} {8}},\
  \bibinfo {pages} {1901192} (\bibinfo {year} {2020})}\BibitemShut {NoStop}%
\bibitem [{\citenamefont {Lu}\ \emph {et~al.}(2017)\citenamefont {Lu},
  \citenamefont {Ge}, \citenamefont {Liu}, \citenamefont {Lu}, \citenamefont
  {Li}, \citenamefont {Lai}, \citenamefont {Zhao}, \citenamefont {Liao},
  \citenamefont {Jia},\ and\ \citenamefont {Sun}}]{lu2017ultrafast}%
  \BibitemOpen
  \bibfield  {author} {\bibinfo {author} {\bibfnamefont {W.}~\bibnamefont
  {Lu}}, \bibinfo {author} {\bibfnamefont {S.}~\bibnamefont {Ge}}, \bibinfo
  {author} {\bibfnamefont {X.}~\bibnamefont {Liu}}, \bibinfo {author}
  {\bibfnamefont {H.}~\bibnamefont {Lu}}, \bibinfo {author} {\bibfnamefont
  {C.}~\bibnamefont {Li}}, \bibinfo {author} {\bibfnamefont {J.}~\bibnamefont
  {Lai}}, \bibinfo {author} {\bibfnamefont {C.}~\bibnamefont {Zhao}}, \bibinfo
  {author} {\bibfnamefont {Z.}~\bibnamefont {Liao}}, \bibinfo {author}
  {\bibfnamefont {S.}~\bibnamefont {Jia}},\ and\ \bibinfo {author}
  {\bibfnamefont {D.}~\bibnamefont {Sun}},\ }\href@noop {} {\bibfield
  {journal} {\bibinfo  {journal} {Physical Review B}\ }\textbf {\bibinfo
  {volume} {95}},\ \bibinfo {pages} {024303} (\bibinfo {year}
  {2017})}\BibitemShut {NoStop}%
\bibitem [{\citenamefont {Ooi}\ \emph {et~al.}(2019)\citenamefont {Ooi},
  \citenamefont {Ang}, \citenamefont {Zhai}, \citenamefont {Tan}, \citenamefont
  {Ang},\ and\ \citenamefont {Ong}}]{ooi2019nonlinear}%
  \BibitemOpen
  \bibfield  {author} {\bibinfo {author} {\bibfnamefont {K.~J.}\ \bibnamefont
  {Ooi}}, \bibinfo {author} {\bibfnamefont {Y.}~\bibnamefont {Ang}}, \bibinfo
  {author} {\bibfnamefont {Q.}~\bibnamefont {Zhai}}, \bibinfo {author}
  {\bibfnamefont {D.~T.}\ \bibnamefont {Tan}}, \bibinfo {author} {\bibfnamefont
  {L.}~\bibnamefont {Ang}},\ and\ \bibinfo {author} {\bibfnamefont
  {C.}~\bibnamefont {Ong}},\ }\href@noop {} {\bibfield  {journal} {\bibinfo
  {journal} {APL Photonics}\ }\textbf {\bibinfo {volume} {4}},\ \bibinfo
  {pages} {034402} (\bibinfo {year} {2019})}\BibitemShut {NoStop}%
\bibitem [{\citenamefont {Zhang}\ \emph {et~al.}(2019)\citenamefont {Zhang},
  \citenamefont {Ooi}, \citenamefont {Chen}, \citenamefont {Ang},\ and\
  \citenamefont {Ang}}]{zhang2019optical}%
  \BibitemOpen
  \bibfield  {author} {\bibinfo {author} {\bibfnamefont {T.}~\bibnamefont
  {Zhang}}, \bibinfo {author} {\bibfnamefont {K.}~\bibnamefont {Ooi}}, \bibinfo
  {author} {\bibfnamefont {W.}~\bibnamefont {Chen}}, \bibinfo {author}
  {\bibfnamefont {L.}~\bibnamefont {Ang}},\ and\ \bibinfo {author}
  {\bibfnamefont {Y.~S.}\ \bibnamefont {Ang}},\ }\href@noop {} {\bibfield
  {journal} {\bibinfo  {journal} {Optics express}\ }\textbf {\bibinfo {volume}
  {27}},\ \bibinfo {pages} {38270} (\bibinfo {year} {2019})}\BibitemShut
  {NoStop}%
\bibitem [{\citenamefont {Ullah}\ \emph {et~al.}(2020)\citenamefont {Ullah},
  \citenamefont {Meng}, \citenamefont {Sun}, \citenamefont {Yang},
  \citenamefont {Wang}, \citenamefont {Wang}, \citenamefont {Wang},
  \citenamefont {Xiu}, \citenamefont {Shi},\ and\ \citenamefont
  {Wang}}]{ullah2020third}%
  \BibitemOpen
  \bibfield  {author} {\bibinfo {author} {\bibfnamefont {K.}~\bibnamefont
  {Ullah}}, \bibinfo {author} {\bibfnamefont {Y.}~\bibnamefont {Meng}},
  \bibinfo {author} {\bibfnamefont {Y.}~\bibnamefont {Sun}}, \bibinfo {author}
  {\bibfnamefont {Y.}~\bibnamefont {Yang}}, \bibinfo {author} {\bibfnamefont
  {X.}~\bibnamefont {Wang}}, \bibinfo {author} {\bibfnamefont {A.}~\bibnamefont
  {Wang}}, \bibinfo {author} {\bibfnamefont {X.}~\bibnamefont {Wang}}, \bibinfo
  {author} {\bibfnamefont {F.}~\bibnamefont {Xiu}}, \bibinfo {author}
  {\bibfnamefont {Y.}~\bibnamefont {Shi}},\ and\ \bibinfo {author}
  {\bibfnamefont {F.}~\bibnamefont {Wang}},\ }\href@noop {} {\bibfield
  {journal} {\bibinfo  {journal} {Applied Physics Letters}\ }\textbf {\bibinfo
  {volume} {117}},\ \bibinfo {pages} {011102} (\bibinfo {year}
  {2020})}\BibitemShut {NoStop}%
\bibitem [{\citenamefont {Lim}\ \emph {et~al.}(2021)\citenamefont {Lim},
  \citenamefont {Ang}, \citenamefont {Ang},\ and\ \citenamefont
  {Wong}}]{lim2021maximal}%
  \BibitemOpen
  \bibfield  {author} {\bibinfo {author} {\bibfnamefont {J.}~\bibnamefont
  {Lim}}, \bibinfo {author} {\bibfnamefont {Y.~S.}\ \bibnamefont {Ang}},
  \bibinfo {author} {\bibfnamefont {L.~K.}\ \bibnamefont {Ang}},\ and\ \bibinfo
  {author} {\bibfnamefont {L.~J.}\ \bibnamefont {Wong}},\ }\href@noop {}
  {\bibfield  {journal} {\bibinfo  {journal} {Communications Physics}\ }\textbf
  {\bibinfo {volume} {4}},\ \bibinfo {pages} {1} (\bibinfo {year}
  {2021})}\BibitemShut {NoStop}%
\bibitem [{\citenamefont {Lim}\ \emph {et~al.}(2020{\natexlab{a}})\citenamefont
  {Lim}, \citenamefont {Ang}, \citenamefont {de~Abajo}, \citenamefont
  {Kaminer}, \citenamefont {Ang},\ and\ \citenamefont
  {Wong}}]{lim2020efficient}%
  \BibitemOpen
  \bibfield  {author} {\bibinfo {author} {\bibfnamefont {J.}~\bibnamefont
  {Lim}}, \bibinfo {author} {\bibfnamefont {Y.~S.}\ \bibnamefont {Ang}},
  \bibinfo {author} {\bibfnamefont {F.~J.~G.}\ \bibnamefont {de~Abajo}},
  \bibinfo {author} {\bibfnamefont {I.}~\bibnamefont {Kaminer}}, \bibinfo
  {author} {\bibfnamefont {L.~K.}\ \bibnamefont {Ang}},\ and\ \bibinfo {author}
  {\bibfnamefont {L.~J.}\ \bibnamefont {Wong}},\ }\href@noop {} {\bibfield
  {journal} {\bibinfo  {journal} {Physical Review Research}\ }\textbf {\bibinfo
  {volume} {2}},\ \bibinfo {pages} {043252} (\bibinfo {year}
  {2020}{\natexlab{a}})}\BibitemShut {NoStop}%
\bibitem [{\citenamefont {Cheng}\ \emph
  {et~al.}(2020{\natexlab{a}})\citenamefont {Cheng}, \citenamefont {Kanda},
  \citenamefont {Ikeda}, \citenamefont {Matsuda}, \citenamefont {Xia},
  \citenamefont {Schumann}, \citenamefont {Stemmer}, \citenamefont {Itatani},
  \citenamefont {Armitage},\ and\ \citenamefont
  {Matsunaga}}]{cheng2020efficient}%
  \BibitemOpen
  \bibfield  {author} {\bibinfo {author} {\bibfnamefont {B.}~\bibnamefont
  {Cheng}}, \bibinfo {author} {\bibfnamefont {N.}~\bibnamefont {Kanda}},
  \bibinfo {author} {\bibfnamefont {T.~N.}\ \bibnamefont {Ikeda}}, \bibinfo
  {author} {\bibfnamefont {T.}~\bibnamefont {Matsuda}}, \bibinfo {author}
  {\bibfnamefont {P.}~\bibnamefont {Xia}}, \bibinfo {author} {\bibfnamefont
  {T.}~\bibnamefont {Schumann}}, \bibinfo {author} {\bibfnamefont
  {S.}~\bibnamefont {Stemmer}}, \bibinfo {author} {\bibfnamefont
  {J.}~\bibnamefont {Itatani}}, \bibinfo {author} {\bibfnamefont
  {N.}~\bibnamefont {Armitage}},\ and\ \bibinfo {author} {\bibfnamefont
  {R.}~\bibnamefont {Matsunaga}},\ }\href@noop {} {\bibfield  {journal}
  {\bibinfo  {journal} {Physical Review Letters}\ }\textbf {\bibinfo {volume}
  {124}},\ \bibinfo {pages} {117402} (\bibinfo {year}
  {2020}{\natexlab{a}})}\BibitemShut {NoStop}%
\bibitem [{\citenamefont {Kovalev}\ \emph {et~al.}(2020)\citenamefont
  {Kovalev}, \citenamefont {Dantas}, \citenamefont {Germanskiy}, \citenamefont
  {Deinert}, \citenamefont {Green}, \citenamefont {Ilyakov}, \citenamefont
  {Awari}, \citenamefont {Chen}, \citenamefont {Bawatna}, \citenamefont {Ling}
  \emph {et~al.}}]{kovalev2020non}%
  \BibitemOpen
  \bibfield  {author} {\bibinfo {author} {\bibfnamefont {S.}~\bibnamefont
  {Kovalev}}, \bibinfo {author} {\bibfnamefont {R.}~\bibnamefont {Dantas}},
  \bibinfo {author} {\bibfnamefont {S.}~\bibnamefont {Germanskiy}}, \bibinfo
  {author} {\bibfnamefont {J.-C.}\ \bibnamefont {Deinert}}, \bibinfo {author}
  {\bibfnamefont {B.}~\bibnamefont {Green}}, \bibinfo {author} {\bibfnamefont
  {I.}~\bibnamefont {Ilyakov}}, \bibinfo {author} {\bibfnamefont
  {N.}~\bibnamefont {Awari}}, \bibinfo {author} {\bibfnamefont
  {M.}~\bibnamefont {Chen}}, \bibinfo {author} {\bibfnamefont {M.}~\bibnamefont
  {Bawatna}}, \bibinfo {author} {\bibfnamefont {J.}~\bibnamefont {Ling}}, \emph
  {et~al.},\ }\href@noop {} {\bibfield  {journal} {\bibinfo  {journal} {Nature
  communications}\ }\textbf {\bibinfo {volume} {11}},\ \bibinfo {pages} {1}
  (\bibinfo {year} {2020})}\BibitemShut {NoStop}%
\bibitem [{\citenamefont {Cheng}\ \emph
  {et~al.}(2020{\natexlab{b}})\citenamefont {Cheng}, \citenamefont {Sipe},\
  and\ \citenamefont {Wu}}]{cheng2020third}%
  \BibitemOpen
  \bibfield  {author} {\bibinfo {author} {\bibfnamefont {J.}~\bibnamefont
  {Cheng}}, \bibinfo {author} {\bibfnamefont {J.}~\bibnamefont {Sipe}},\ and\
  \bibinfo {author} {\bibfnamefont {S.}~\bibnamefont {Wu}},\ }\href@noop {}
  {\bibfield  {journal} {\bibinfo  {journal} {ACS Photonics}\ }\textbf
  {\bibinfo {volume} {7}},\ \bibinfo {pages} {2515} (\bibinfo {year}
  {2020}{\natexlab{b}})}\BibitemShut {NoStop}%
\bibitem [{\citenamefont {Lee}\ \emph {et~al.}(2020)\citenamefont {Lee},
  \citenamefont {Yap}, \citenamefont {Tai}, \citenamefont {Xu}, \citenamefont
  {Zhang},\ and\ \citenamefont {Gong}}]{lee2020enhanced}%
  \BibitemOpen
  \bibfield  {author} {\bibinfo {author} {\bibfnamefont {C.~H.}\ \bibnamefont
  {Lee}}, \bibinfo {author} {\bibfnamefont {H.~H.}\ \bibnamefont {Yap}},
  \bibinfo {author} {\bibfnamefont {T.}~\bibnamefont {Tai}}, \bibinfo {author}
  {\bibfnamefont {G.}~\bibnamefont {Xu}}, \bibinfo {author} {\bibfnamefont
  {X.}~\bibnamefont {Zhang}},\ and\ \bibinfo {author} {\bibfnamefont
  {J.}~\bibnamefont {Gong}},\ }\href@noop {} {\bibfield  {journal} {\bibinfo
  {journal} {Physical Review B}\ }\textbf {\bibinfo {volume} {102}},\ \bibinfo
  {pages} {035138} (\bibinfo {year} {2020})}\BibitemShut {NoStop}%
\bibitem [{\citenamefont {Tai}\ and\ \citenamefont
  {Lee}(2021)}]{tai2021anisotropic}%
  \BibitemOpen
  \bibfield  {author} {\bibinfo {author} {\bibfnamefont {T.}~\bibnamefont
  {Tai}}\ and\ \bibinfo {author} {\bibfnamefont {C.~H.}\ \bibnamefont {Lee}},\
  }\href@noop {} {\bibfield  {journal} {\bibinfo  {journal} {Physical Review
  B}\ }\textbf {\bibinfo {volume} {103}},\ \bibinfo {pages} {195125} (\bibinfo
  {year} {2021})}\BibitemShut {NoStop}%
\bibitem [{\citenamefont {Zhang}\ \emph {et~al.}(2011)\citenamefont {Zhang},
  \citenamefont {Yu}, \citenamefont {Feng}, \citenamefont {Yao}, \citenamefont
  {Weng}, \citenamefont {Dai},\ and\ \citenamefont
  {Fang}}]{zhang2011topological}%
  \BibitemOpen
  \bibfield  {author} {\bibinfo {author} {\bibfnamefont {W.}~\bibnamefont
  {Zhang}}, \bibinfo {author} {\bibfnamefont {R.}~\bibnamefont {Yu}}, \bibinfo
  {author} {\bibfnamefont {W.}~\bibnamefont {Feng}}, \bibinfo {author}
  {\bibfnamefont {Y.}~\bibnamefont {Yao}}, \bibinfo {author} {\bibfnamefont
  {H.}~\bibnamefont {Weng}}, \bibinfo {author} {\bibfnamefont {X.}~\bibnamefont
  {Dai}},\ and\ \bibinfo {author} {\bibfnamefont {Z.}~\bibnamefont {Fang}},\
  }\href@noop {} {\bibfield  {journal} {\bibinfo  {journal} {Physical review
  letters}\ }\textbf {\bibinfo {volume} {106}},\ \bibinfo {pages} {156808}
  (\bibinfo {year} {2011})}\BibitemShut {NoStop}%
\bibitem [{\citenamefont {Wang}\ \emph {et~al.}(2021)\citenamefont {Wang},
  \citenamefont {Xiang}, \citenamefont {Ye}, \citenamefont {Zheng},
  \citenamefont {Yu},\ and\ \citenamefont {Liao}}]{wang2021surface}%
  \BibitemOpen
  \bibfield  {author} {\bibinfo {author} {\bibfnamefont {A.-Q.}\ \bibnamefont
  {Wang}}, \bibinfo {author} {\bibfnamefont {P.-Z.}\ \bibnamefont {Xiang}},
  \bibinfo {author} {\bibfnamefont {X.-G.}\ \bibnamefont {Ye}}, \bibinfo
  {author} {\bibfnamefont {W.-Z.}\ \bibnamefont {Zheng}}, \bibinfo {author}
  {\bibfnamefont {D.}~\bibnamefont {Yu}},\ and\ \bibinfo {author}
  {\bibfnamefont {Z.-M.}\ \bibnamefont {Liao}},\ }\href@noop {} {\bibfield
  {journal} {\bibinfo  {journal} {Nano Letters}\ }\textbf {\bibinfo {volume}
  {21}},\ \bibinfo {pages} {2026} (\bibinfo {year} {2021})}\BibitemShut
  {NoStop}%
\bibitem [{\citenamefont {Schumann}\ \emph {et~al.}(2018)\citenamefont
  {Schumann}, \citenamefont {Galletti}, \citenamefont {Kealhofer},
  \citenamefont {Kim}, \citenamefont {Goyal},\ and\ \citenamefont
  {Stemmer}}]{schumann2018observation}%
  \BibitemOpen
  \bibfield  {author} {\bibinfo {author} {\bibfnamefont {T.}~\bibnamefont
  {Schumann}}, \bibinfo {author} {\bibfnamefont {L.}~\bibnamefont {Galletti}},
  \bibinfo {author} {\bibfnamefont {D.~A.}\ \bibnamefont {Kealhofer}}, \bibinfo
  {author} {\bibfnamefont {H.}~\bibnamefont {Kim}}, \bibinfo {author}
  {\bibfnamefont {M.}~\bibnamefont {Goyal}},\ and\ \bibinfo {author}
  {\bibfnamefont {S.}~\bibnamefont {Stemmer}},\ }\href@noop {} {\bibfield
  {journal} {\bibinfo  {journal} {Physical review letters}\ }\textbf {\bibinfo
  {volume} {120}},\ \bibinfo {pages} {016801} (\bibinfo {year}
  {2018})}\BibitemShut {NoStop}%
\bibitem [{\citenamefont {Goyal}\ \emph {et~al.}(2018)\citenamefont {Goyal},
  \citenamefont {Galletti}, \citenamefont {Salmani-Rezaie}, \citenamefont
  {Schumann}, \citenamefont {Kealhofer},\ and\ \citenamefont
  {Stemmer}}]{goyal2018thickness}%
  \BibitemOpen
  \bibfield  {author} {\bibinfo {author} {\bibfnamefont {M.}~\bibnamefont
  {Goyal}}, \bibinfo {author} {\bibfnamefont {L.}~\bibnamefont {Galletti}},
  \bibinfo {author} {\bibfnamefont {S.}~\bibnamefont {Salmani-Rezaie}},
  \bibinfo {author} {\bibfnamefont {T.}~\bibnamefont {Schumann}}, \bibinfo
  {author} {\bibfnamefont {D.~A.}\ \bibnamefont {Kealhofer}},\ and\ \bibinfo
  {author} {\bibfnamefont {S.}~\bibnamefont {Stemmer}},\ }\href@noop {}
  {\bibfield  {journal} {\bibinfo  {journal} {APL Materials}\ }\textbf
  {\bibinfo {volume} {6}},\ \bibinfo {pages} {026105} (\bibinfo {year}
  {2018})}\BibitemShut {NoStop}%
\bibitem [{\citenamefont {Zhang}\ \emph {et~al.}(2017)\citenamefont {Zhang},
  \citenamefont {Narayan}, \citenamefont {Lu}, \citenamefont {Zhang},
  \citenamefont {Zhang}, \citenamefont {Ni}, \citenamefont {Yuan},
  \citenamefont {Liu}, \citenamefont {Park}, \citenamefont {Zhang} \emph
  {et~al.}}]{zhang2017evolution}%
  \BibitemOpen
  \bibfield  {author} {\bibinfo {author} {\bibfnamefont {C.}~\bibnamefont
  {Zhang}}, \bibinfo {author} {\bibfnamefont {A.}~\bibnamefont {Narayan}},
  \bibinfo {author} {\bibfnamefont {S.}~\bibnamefont {Lu}}, \bibinfo {author}
  {\bibfnamefont {J.}~\bibnamefont {Zhang}}, \bibinfo {author} {\bibfnamefont
  {H.}~\bibnamefont {Zhang}}, \bibinfo {author} {\bibfnamefont
  {Z.}~\bibnamefont {Ni}}, \bibinfo {author} {\bibfnamefont {X.}~\bibnamefont
  {Yuan}}, \bibinfo {author} {\bibfnamefont {Y.}~\bibnamefont {Liu}}, \bibinfo
  {author} {\bibfnamefont {J.-H.}\ \bibnamefont {Park}}, \bibinfo {author}
  {\bibfnamefont {E.}~\bibnamefont {Zhang}}, \emph {et~al.},\ }\href@noop {}
  {\bibfield  {journal} {\bibinfo  {journal} {Nature communications}\ }\textbf
  {\bibinfo {volume} {8}},\ \bibinfo {pages} {1} (\bibinfo {year}
  {2017})}\BibitemShut {NoStop}%
\bibitem [{\citenamefont {Wang}\ \emph
  {et~al.}(2020{\natexlab{a}})\citenamefont {Wang}, \citenamefont {Ye},
  \citenamefont {Yu},\ and\ \citenamefont {Liao}}]{wang2020topological}%
  \BibitemOpen
  \bibfield  {author} {\bibinfo {author} {\bibfnamefont {A.-Q.}\ \bibnamefont
  {Wang}}, \bibinfo {author} {\bibfnamefont {X.-G.}\ \bibnamefont {Ye}},
  \bibinfo {author} {\bibfnamefont {D.-P.}\ \bibnamefont {Yu}},\ and\ \bibinfo
  {author} {\bibfnamefont {Z.-M.}\ \bibnamefont {Liao}},\ }\href@noop {}
  {\bibfield  {journal} {\bibinfo  {journal} {ACS nano}\ }\textbf {\bibinfo
  {volume} {14}},\ \bibinfo {pages} {3755} (\bibinfo {year}
  {2020}{\natexlab{a}})}\BibitemShut {NoStop}%
\bibitem [{\citenamefont {Liu}\ \emph {et~al.}(2020)\citenamefont {Liu},
  \citenamefont {Xia}, \citenamefont {Xiao}, \citenamefont {Garcia~de Abajo},\
  and\ \citenamefont {Sun}}]{liu2020semimetals}%
  \BibitemOpen
  \bibfield  {author} {\bibinfo {author} {\bibfnamefont {J.}~\bibnamefont
  {Liu}}, \bibinfo {author} {\bibfnamefont {F.}~\bibnamefont {Xia}}, \bibinfo
  {author} {\bibfnamefont {D.}~\bibnamefont {Xiao}}, \bibinfo {author}
  {\bibfnamefont {F.~J.}\ \bibnamefont {Garcia~de Abajo}},\ and\ \bibinfo
  {author} {\bibfnamefont {D.}~\bibnamefont {Sun}},\ }\href@noop {} {\bibfield
  {journal} {\bibinfo  {journal} {Nature materials}\ }\textbf {\bibinfo
  {volume} {19}},\ \bibinfo {pages} {830} (\bibinfo {year} {2020})}\BibitemShut
  {NoStop}%
\bibitem [{\citenamefont {Collins}\ \emph {et~al.}(2018)\citenamefont
  {Collins}, \citenamefont {Tadich}, \citenamefont {Wu}, \citenamefont {Gomes},
  \citenamefont {Rodrigues}, \citenamefont {Liu}, \citenamefont {Hellerstedt},
  \citenamefont {Ryu}, \citenamefont {Tang}, \citenamefont {Mo} \emph
  {et~al.}}]{collins2018electric}%
  \BibitemOpen
  \bibfield  {author} {\bibinfo {author} {\bibfnamefont {J.~L.}\ \bibnamefont
  {Collins}}, \bibinfo {author} {\bibfnamefont {A.}~\bibnamefont {Tadich}},
  \bibinfo {author} {\bibfnamefont {W.}~\bibnamefont {Wu}}, \bibinfo {author}
  {\bibfnamefont {L.~C.}\ \bibnamefont {Gomes}}, \bibinfo {author}
  {\bibfnamefont {J.~N.}\ \bibnamefont {Rodrigues}}, \bibinfo {author}
  {\bibfnamefont {C.}~\bibnamefont {Liu}}, \bibinfo {author} {\bibfnamefont
  {J.}~\bibnamefont {Hellerstedt}}, \bibinfo {author} {\bibfnamefont
  {H.}~\bibnamefont {Ryu}}, \bibinfo {author} {\bibfnamefont {S.}~\bibnamefont
  {Tang}}, \bibinfo {author} {\bibfnamefont {S.-K.}\ \bibnamefont {Mo}}, \emph
  {et~al.},\ }\href@noop {} {\bibfield  {journal} {\bibinfo  {journal}
  {Nature}\ }\textbf {\bibinfo {volume} {564}},\ \bibinfo {pages} {390}
  (\bibinfo {year} {2018})}\BibitemShut {NoStop}%
\bibitem [{\citenamefont {Suess}\ \emph {et~al.}(2018)\citenamefont {Suess},
  \citenamefont {Bachleitner-Hofmann}, \citenamefont {Satz}, \citenamefont
  {Weitensfelder}, \citenamefont {Vogler}, \citenamefont {Bruckner},
  \citenamefont {Abert}, \citenamefont {Pr{\"u}gl}, \citenamefont {Zimmer},
  \citenamefont {Huber} \emph {et~al.}}]{suess2018topologically}%
  \BibitemOpen
  \bibfield  {author} {\bibinfo {author} {\bibfnamefont {D.}~\bibnamefont
  {Suess}}, \bibinfo {author} {\bibfnamefont {A.}~\bibnamefont
  {Bachleitner-Hofmann}}, \bibinfo {author} {\bibfnamefont {A.}~\bibnamefont
  {Satz}}, \bibinfo {author} {\bibfnamefont {H.}~\bibnamefont {Weitensfelder}},
  \bibinfo {author} {\bibfnamefont {C.}~\bibnamefont {Vogler}}, \bibinfo
  {author} {\bibfnamefont {F.}~\bibnamefont {Bruckner}}, \bibinfo {author}
  {\bibfnamefont {C.}~\bibnamefont {Abert}}, \bibinfo {author} {\bibfnamefont
  {K.}~\bibnamefont {Pr{\"u}gl}}, \bibinfo {author} {\bibfnamefont
  {J.}~\bibnamefont {Zimmer}}, \bibinfo {author} {\bibfnamefont
  {C.}~\bibnamefont {Huber}}, \emph {et~al.},\ }\href@noop {} {\bibfield
  {journal} {\bibinfo  {journal} {Nature Electronics}\ }\textbf {\bibinfo
  {volume} {1}},\ \bibinfo {pages} {362} (\bibinfo {year} {2018})}\BibitemShut
  {NoStop}%
\bibitem [{\citenamefont {Wu}\ \emph {et~al.}(2018)\citenamefont {Wu},
  \citenamefont {Zhang}, \citenamefont {Li}, \citenamefont {Zhang},
  \citenamefont {Liu}, \citenamefont {Liao},\ and\ \citenamefont
  {Yu}}]{wu2018dirac}%
  \BibitemOpen
  \bibfield  {author} {\bibinfo {author} {\bibfnamefont {Y.-F.}\ \bibnamefont
  {Wu}}, \bibinfo {author} {\bibfnamefont {L.}~\bibnamefont {Zhang}}, \bibinfo
  {author} {\bibfnamefont {C.-Z.}\ \bibnamefont {Li}}, \bibinfo {author}
  {\bibfnamefont {Z.-S.}\ \bibnamefont {Zhang}}, \bibinfo {author}
  {\bibfnamefont {S.}~\bibnamefont {Liu}}, \bibinfo {author} {\bibfnamefont
  {Z.-M.}\ \bibnamefont {Liao}},\ and\ \bibinfo {author} {\bibfnamefont
  {D.}~\bibnamefont {Yu}},\ }\href@noop {} {\bibfield  {journal} {\bibinfo
  {journal} {Advanced Materials}\ }\textbf {\bibinfo {volume} {30}},\ \bibinfo
  {pages} {1707547} (\bibinfo {year} {2018})}\BibitemShut {NoStop}%
\bibitem [{\citenamefont {Fu}\ \emph {et~al.}(2020)\citenamefont {Fu},
  \citenamefont {Sun},\ and\ \citenamefont {Felser}}]{fu2020topological}%
  \BibitemOpen
  \bibfield  {author} {\bibinfo {author} {\bibfnamefont {C.}~\bibnamefont
  {Fu}}, \bibinfo {author} {\bibfnamefont {Y.}~\bibnamefont {Sun}},\ and\
  \bibinfo {author} {\bibfnamefont {C.}~\bibnamefont {Felser}},\ }\href@noop {}
  {\bibfield  {journal} {\bibinfo  {journal} {APL Materials}\ }\textbf
  {\bibinfo {volume} {8}},\ \bibinfo {pages} {040913} (\bibinfo {year}
  {2020})}\BibitemShut {NoStop}%
\bibitem [{\citenamefont {Cao}\ \emph {et~al.}(2020)\citenamefont {Cao},
  \citenamefont {Zhou}, \citenamefont {Wu}, \citenamefont {Yang}, \citenamefont
  {Yang}, \citenamefont {Ang},\ and\ \citenamefont {Ang}}]{cao2020electrical}%
  \BibitemOpen
  \bibfield  {author} {\bibinfo {author} {\bibfnamefont {L.}~\bibnamefont
  {Cao}}, \bibinfo {author} {\bibfnamefont {G.}~\bibnamefont {Zhou}}, \bibinfo
  {author} {\bibfnamefont {Q.}~\bibnamefont {Wu}}, \bibinfo {author}
  {\bibfnamefont {S.~A.}\ \bibnamefont {Yang}}, \bibinfo {author}
  {\bibfnamefont {H.~Y.}\ \bibnamefont {Yang}}, \bibinfo {author}
  {\bibfnamefont {Y.~S.}\ \bibnamefont {Ang}},\ and\ \bibinfo {author}
  {\bibfnamefont {L.}~\bibnamefont {Ang}},\ }\href@noop {} {\bibfield
  {journal} {\bibinfo  {journal} {Physical Review Applied}\ }\textbf {\bibinfo
  {volume} {13}},\ \bibinfo {pages} {054030} (\bibinfo {year}
  {2020})}\BibitemShut {NoStop}%
\bibitem [{\citenamefont {Dai}\ \emph {et~al.}(2021)\citenamefont {Dai},
  \citenamefont {Manjappa}, \citenamefont {Yang}, \citenamefont {Tan},
  \citenamefont {Qiang}, \citenamefont {Han}, \citenamefont {Wong},
  \citenamefont {Xiu}, \citenamefont {Liu},\ and\ \citenamefont
  {Singh}}]{dai2021high}%
  \BibitemOpen
  \bibfield  {author} {\bibinfo {author} {\bibfnamefont {Z.}~\bibnamefont
  {Dai}}, \bibinfo {author} {\bibfnamefont {M.}~\bibnamefont {Manjappa}},
  \bibinfo {author} {\bibfnamefont {Y.}~\bibnamefont {Yang}}, \bibinfo {author}
  {\bibfnamefont {T.~C.~W.}\ \bibnamefont {Tan}}, \bibinfo {author}
  {\bibfnamefont {B.}~\bibnamefont {Qiang}}, \bibinfo {author} {\bibfnamefont
  {S.}~\bibnamefont {Han}}, \bibinfo {author} {\bibfnamefont {L.~J.}\
  \bibnamefont {Wong}}, \bibinfo {author} {\bibfnamefont {F.}~\bibnamefont
  {Xiu}}, \bibinfo {author} {\bibfnamefont {W.}~\bibnamefont {Liu}},\ and\
  \bibinfo {author} {\bibfnamefont {R.}~\bibnamefont {Singh}},\ }\href@noop {}
  {\bibfield  {journal} {\bibinfo  {journal} {Advanced Functional Materials}\
  }\textbf {\bibinfo {volume} {31}},\ \bibinfo {pages} {2011011} (\bibinfo
  {year} {2021})}\BibitemShut {NoStop}%
\bibitem [{\citenamefont {Zhu}\ \emph {et~al.}(2017)\citenamefont {Zhu},
  \citenamefont {Wang}, \citenamefont {Meng}, \citenamefont {Yuan},
  \citenamefont {Xiu}, \citenamefont {Luo}, \citenamefont {Wang}, \citenamefont
  {Li}, \citenamefont {Lv}, \citenamefont {He} \emph {et~al.}}]{zhu2017robust}%
  \BibitemOpen
  \bibfield  {author} {\bibinfo {author} {\bibfnamefont {C.}~\bibnamefont
  {Zhu}}, \bibinfo {author} {\bibfnamefont {F.}~\bibnamefont {Wang}}, \bibinfo
  {author} {\bibfnamefont {Y.}~\bibnamefont {Meng}}, \bibinfo {author}
  {\bibfnamefont {X.}~\bibnamefont {Yuan}}, \bibinfo {author} {\bibfnamefont
  {F.}~\bibnamefont {Xiu}}, \bibinfo {author} {\bibfnamefont {H.}~\bibnamefont
  {Luo}}, \bibinfo {author} {\bibfnamefont {Y.}~\bibnamefont {Wang}}, \bibinfo
  {author} {\bibfnamefont {J.}~\bibnamefont {Li}}, \bibinfo {author}
  {\bibfnamefont {X.}~\bibnamefont {Lv}}, \bibinfo {author} {\bibfnamefont
  {L.}~\bibnamefont {He}}, \emph {et~al.},\ }\href@noop {} {\bibfield
  {journal} {\bibinfo  {journal} {Nature communications}\ }\textbf {\bibinfo
  {volume} {8}},\ \bibinfo {pages} {1} (\bibinfo {year} {2017})}\BibitemShut
  {NoStop}%
\bibitem [{\citenamefont {Meng}\ \emph {et~al.}(2019)\citenamefont {Meng},
  \citenamefont {Shang}, \citenamefont {Xue}, \citenamefont {Tang},
  \citenamefont {Xia}, \citenamefont {Zhai}, \citenamefont {Liu}, \citenamefont
  {Chen}, \citenamefont {Li},\ and\ \citenamefont
  {Wang}}]{meng2019bidirectional}%
  \BibitemOpen
  \bibfield  {author} {\bibinfo {author} {\bibfnamefont {H.}~\bibnamefont
  {Meng}}, \bibinfo {author} {\bibfnamefont {X.}~\bibnamefont {Shang}},
  \bibinfo {author} {\bibfnamefont {X.}~\bibnamefont {Xue}}, \bibinfo {author}
  {\bibfnamefont {K.}~\bibnamefont {Tang}}, \bibinfo {author} {\bibfnamefont
  {S.}~\bibnamefont {Xia}}, \bibinfo {author} {\bibfnamefont {X.}~\bibnamefont
  {Zhai}}, \bibinfo {author} {\bibfnamefont {Z.}~\bibnamefont {Liu}}, \bibinfo
  {author} {\bibfnamefont {J.}~\bibnamefont {Chen}}, \bibinfo {author}
  {\bibfnamefont {H.}~\bibnamefont {Li}},\ and\ \bibinfo {author}
  {\bibfnamefont {L.}~\bibnamefont {Wang}},\ }\href@noop {} {\bibfield
  {journal} {\bibinfo  {journal} {Optics express}\ }\textbf {\bibinfo {volume}
  {27}},\ \bibinfo {pages} {31062} (\bibinfo {year} {2019})}\BibitemShut
  {NoStop}%
\bibitem [{\citenamefont {Li}\ \emph {et~al.}(2021)\citenamefont {Li},
  \citenamefont {Yi}, \citenamefont {Xu}, \citenamefont {Yang}, \citenamefont
  {Yi}, \citenamefont {Chen}, \citenamefont {Yi}, \citenamefont {Zhang},\ and\
  \citenamefont {Wu}}]{li2021multi}%
  \BibitemOpen
  \bibfield  {author} {\bibinfo {author} {\bibfnamefont {Z.}~\bibnamefont
  {Li}}, \bibinfo {author} {\bibfnamefont {Y.}~\bibnamefont {Yi}}, \bibinfo
  {author} {\bibfnamefont {D.}~\bibnamefont {Xu}}, \bibinfo {author}
  {\bibfnamefont {H.}~\bibnamefont {Yang}}, \bibinfo {author} {\bibfnamefont
  {Z.}~\bibnamefont {Yi}}, \bibinfo {author} {\bibfnamefont {X.}~\bibnamefont
  {Chen}}, \bibinfo {author} {\bibfnamefont {Y.}~\bibnamefont {Yi}}, \bibinfo
  {author} {\bibfnamefont {J.}~\bibnamefont {Zhang}},\ and\ \bibinfo {author}
  {\bibfnamefont {P.}~\bibnamefont {Wu}},\ }\href@noop {} {\bibfield  {journal}
  {\bibinfo  {journal} {Chinese Physics B}\ } (\bibinfo {year}
  {2021})}\BibitemShut {NoStop}%
\bibitem [{\citenamefont {Wang}\ \emph {et~al.}(2017)\citenamefont {Wang},
  \citenamefont {Li}, \citenamefont {Ge}, \citenamefont {Li}, \citenamefont
  {Lu}, \citenamefont {Lai}, \citenamefont {Liu}, \citenamefont {Ma},
  \citenamefont {Yu}, \citenamefont {Liao} \emph {et~al.}}]{wang2017ultrafast}%
  \BibitemOpen
  \bibfield  {author} {\bibinfo {author} {\bibfnamefont {Q.}~\bibnamefont
  {Wang}}, \bibinfo {author} {\bibfnamefont {C.-Z.}\ \bibnamefont {Li}},
  \bibinfo {author} {\bibfnamefont {S.}~\bibnamefont {Ge}}, \bibinfo {author}
  {\bibfnamefont {J.-G.}\ \bibnamefont {Li}}, \bibinfo {author} {\bibfnamefont
  {W.}~\bibnamefont {Lu}}, \bibinfo {author} {\bibfnamefont {J.}~\bibnamefont
  {Lai}}, \bibinfo {author} {\bibfnamefont {X.}~\bibnamefont {Liu}}, \bibinfo
  {author} {\bibfnamefont {J.}~\bibnamefont {Ma}}, \bibinfo {author}
  {\bibfnamefont {D.-P.}\ \bibnamefont {Yu}}, \bibinfo {author} {\bibfnamefont
  {Z.-M.}\ \bibnamefont {Liao}}, \emph {et~al.},\ }\href@noop {} {\bibfield
  {journal} {\bibinfo  {journal} {Nano Letters}\ }\textbf {\bibinfo {volume}
  {17}},\ \bibinfo {pages} {834} (\bibinfo {year} {2017})}\BibitemShut
  {NoStop}%
\bibitem [{\citenamefont {Yang}\ \emph
  {et~al.}(2018{\natexlab{a}})\citenamefont {Yang}, \citenamefont {Zhang},
  \citenamefont {Deng}, \citenamefont {Chu}, \citenamefont {Jiang},
  \citenamefont {Meng}, \citenamefont {Wang}, \citenamefont {Zhang},
  \citenamefont {Yin},\ and\ \citenamefont {You}}]{yang2018efficient}%
  \BibitemOpen
  \bibfield  {author} {\bibinfo {author} {\bibfnamefont {X.}~\bibnamefont
  {Yang}}, \bibinfo {author} {\bibfnamefont {X.}~\bibnamefont {Zhang}},
  \bibinfo {author} {\bibfnamefont {J.}~\bibnamefont {Deng}}, \bibinfo {author}
  {\bibfnamefont {Z.}~\bibnamefont {Chu}}, \bibinfo {author} {\bibfnamefont
  {Q.}~\bibnamefont {Jiang}}, \bibinfo {author} {\bibfnamefont
  {J.}~\bibnamefont {Meng}}, \bibinfo {author} {\bibfnamefont {P.}~\bibnamefont
  {Wang}}, \bibinfo {author} {\bibfnamefont {L.}~\bibnamefont {Zhang}},
  \bibinfo {author} {\bibfnamefont {Z.}~\bibnamefont {Yin}},\ and\ \bibinfo
  {author} {\bibfnamefont {J.}~\bibnamefont {You}},\ }\href@noop {} {\bibfield
  {journal} {\bibinfo  {journal} {Nature communications}\ }\textbf {\bibinfo
  {volume} {9}},\ \bibinfo {pages} {1} (\bibinfo {year}
  {2018}{\natexlab{a}})}\BibitemShut {NoStop}%
\bibitem [{\citenamefont {Yang}\ \emph
  {et~al.}(2018{\natexlab{b}})\citenamefont {Yang}, \citenamefont {Wang},
  \citenamefont {Han}, \citenamefont {Ling}, \citenamefont {Ji}, \citenamefont
  {Kong}, \citenamefont {Liu}, \citenamefont {Huang}, \citenamefont {Gou},
  \citenamefont {Liu} \emph {et~al.}}]{yang2018enhanced}%
  \BibitemOpen
  \bibfield  {author} {\bibinfo {author} {\bibfnamefont {M.}~\bibnamefont
  {Yang}}, \bibinfo {author} {\bibfnamefont {J.}~\bibnamefont {Wang}}, \bibinfo
  {author} {\bibfnamefont {J.}~\bibnamefont {Han}}, \bibinfo {author}
  {\bibfnamefont {J.}~\bibnamefont {Ling}}, \bibinfo {author} {\bibfnamefont
  {C.}~\bibnamefont {Ji}}, \bibinfo {author} {\bibfnamefont {X.}~\bibnamefont
  {Kong}}, \bibinfo {author} {\bibfnamefont {X.}~\bibnamefont {Liu}}, \bibinfo
  {author} {\bibfnamefont {Z.}~\bibnamefont {Huang}}, \bibinfo {author}
  {\bibfnamefont {J.}~\bibnamefont {Gou}}, \bibinfo {author} {\bibfnamefont
  {Z.}~\bibnamefont {Liu}}, \emph {et~al.},\ }\href@noop {} {\bibfield
  {journal} {\bibinfo  {journal} {ACS Photonics}\ }\textbf {\bibinfo {volume}
  {5}},\ \bibinfo {pages} {3438} (\bibinfo {year}
  {2018}{\natexlab{b}})}\BibitemShut {NoStop}%
\bibitem [{\citenamefont {Jia}\ \emph {et~al.}(2021)\citenamefont {Jia},
  \citenamefont {Huang}, \citenamefont {Zhou}, \citenamefont {Wang},
  \citenamefont {Zhang},\ and\ \citenamefont {Miao}}]{jia2021temperature}%
  \BibitemOpen
  \bibfield  {author} {\bibinfo {author} {\bibfnamefont {G.}~\bibnamefont
  {Jia}}, \bibinfo {author} {\bibfnamefont {Z.}~\bibnamefont {Huang}}, \bibinfo
  {author} {\bibfnamefont {Y.}~\bibnamefont {Zhou}}, \bibinfo {author}
  {\bibfnamefont {H.}~\bibnamefont {Wang}}, \bibinfo {author} {\bibfnamefont
  {Y.}~\bibnamefont {Zhang}},\ and\ \bibinfo {author} {\bibfnamefont
  {X.}~\bibnamefont {Miao}},\ }\href@noop {} {\bibfield  {journal} {\bibinfo
  {journal} {Physical Chemistry Chemical Physics}\ }\textbf {\bibinfo {volume}
  {23}},\ \bibinfo {pages} {13128} (\bibinfo {year} {2021})}\BibitemShut
  {NoStop}%
\bibitem [{\citenamefont {Dai}\ \emph {et~al.}(2019)\citenamefont {Dai},
  \citenamefont {Zhang}, \citenamefont {O’Hara},\ and\ \citenamefont
  {Zhang}}]{dai2019controllable}%
  \BibitemOpen
  \bibfield  {author} {\bibinfo {author} {\bibfnamefont {L.}~\bibnamefont
  {Dai}}, \bibinfo {author} {\bibfnamefont {Y.}~\bibnamefont {Zhang}}, \bibinfo
  {author} {\bibfnamefont {J.~F.}\ \bibnamefont {O’Hara}},\ and\ \bibinfo
  {author} {\bibfnamefont {H.}~\bibnamefont {Zhang}},\ }\href@noop {}
  {\bibfield  {journal} {\bibinfo  {journal} {Optics express}\ }\textbf
  {\bibinfo {volume} {27}},\ \bibinfo {pages} {35784} (\bibinfo {year}
  {2019})}\BibitemShut {NoStop}%
\bibitem [{\citenamefont {Meng}\ \emph {et~al.}(2018)\citenamefont {Meng},
  \citenamefont {Zhu}, \citenamefont {Li}, \citenamefont {Yuan}, \citenamefont
  {Xiu}, \citenamefont {Shi}, \citenamefont {Xu},\ and\ \citenamefont
  {Wang}}]{meng2018three}%
  \BibitemOpen
  \bibfield  {author} {\bibinfo {author} {\bibfnamefont {Y.}~\bibnamefont
  {Meng}}, \bibinfo {author} {\bibfnamefont {C.}~\bibnamefont {Zhu}}, \bibinfo
  {author} {\bibfnamefont {Y.}~\bibnamefont {Li}}, \bibinfo {author}
  {\bibfnamefont {X.}~\bibnamefont {Yuan}}, \bibinfo {author} {\bibfnamefont
  {F.}~\bibnamefont {Xiu}}, \bibinfo {author} {\bibfnamefont {Y.}~\bibnamefont
  {Shi}}, \bibinfo {author} {\bibfnamefont {Y.}~\bibnamefont {Xu}},\ and\
  \bibinfo {author} {\bibfnamefont {F.}~\bibnamefont {Wang}},\ }\href@noop {}
  {\bibfield  {journal} {\bibinfo  {journal} {Optics letters}\ }\textbf
  {\bibinfo {volume} {43}},\ \bibinfo {pages} {1503} (\bibinfo {year}
  {2018})}\BibitemShut {NoStop}%
\bibitem [{\citenamefont {Xiong}\ \emph {et~al.}(2020)\citenamefont {Xiong},
  \citenamefont {Ji}, \citenamefont {Bashir},\ and\ \citenamefont
  {Yang}}]{xiong2020dual}%
  \BibitemOpen
  \bibfield  {author} {\bibinfo {author} {\bibfnamefont {H.}~\bibnamefont
  {Xiong}}, \bibinfo {author} {\bibfnamefont {Q.}~\bibnamefont {Ji}}, \bibinfo
  {author} {\bibfnamefont {T.}~\bibnamefont {Bashir}},\ and\ \bibinfo {author}
  {\bibfnamefont {F.}~\bibnamefont {Yang}},\ }\href@noop {} {\bibfield
  {journal} {\bibinfo  {journal} {Optics express}\ }\textbf {\bibinfo {volume}
  {28}},\ \bibinfo {pages} {13884} (\bibinfo {year} {2020})}\BibitemShut
  {NoStop}%
\bibitem [{\citenamefont {Lim}\ \emph {et~al.}(2020{\natexlab{b}})\citenamefont
  {Lim}, \citenamefont {Ooi}, \citenamefont {Zhang}, \citenamefont {Ang},\ and\
  \citenamefont {Ang}}]{lim2020broadband}%
  \BibitemOpen
  \bibfield  {author} {\bibinfo {author} {\bibfnamefont {J.}~\bibnamefont
  {Lim}}, \bibinfo {author} {\bibfnamefont {K.}~\bibnamefont {Ooi}}, \bibinfo
  {author} {\bibfnamefont {C.}~\bibnamefont {Zhang}}, \bibinfo {author}
  {\bibfnamefont {L.}~\bibnamefont {Ang}},\ and\ \bibinfo {author}
  {\bibfnamefont {Y.~S.}\ \bibnamefont {Ang}},\ }\href@noop {} {\bibfield
  {journal} {\bibinfo  {journal} {Chinese Physics B}\ }\textbf {\bibinfo
  {volume} {29}},\ \bibinfo {pages} {077802} (\bibinfo {year}
  {2020}{\natexlab{b}})}\BibitemShut {NoStop}%
\bibitem [{\citenamefont {Hao}\ \emph {et~al.}(2007)\citenamefont {Hao},
  \citenamefont {Yuan}, \citenamefont {Ran}, \citenamefont {Jiang},
  \citenamefont {Kong}, \citenamefont {Chan},\ and\ \citenamefont
  {Zhou}}]{hao2007manipulating}%
  \BibitemOpen
  \bibfield  {author} {\bibinfo {author} {\bibfnamefont {J.}~\bibnamefont
  {Hao}}, \bibinfo {author} {\bibfnamefont {Y.}~\bibnamefont {Yuan}}, \bibinfo
  {author} {\bibfnamefont {L.}~\bibnamefont {Ran}}, \bibinfo {author}
  {\bibfnamefont {T.}~\bibnamefont {Jiang}}, \bibinfo {author} {\bibfnamefont
  {J.~A.}\ \bibnamefont {Kong}}, \bibinfo {author} {\bibfnamefont
  {C.}~\bibnamefont {Chan}},\ and\ \bibinfo {author} {\bibfnamefont
  {L.}~\bibnamefont {Zhou}},\ }\href@noop {} {\bibfield  {journal} {\bibinfo
  {journal} {Physical review letters}\ }\textbf {\bibinfo {volume} {99}},\
  \bibinfo {pages} {063908} (\bibinfo {year} {2007})}\BibitemShut {NoStop}%
\bibitem [{\citenamefont {Zhao}\ and\ \citenamefont
  {Alu}(2013)}]{zhao2013tailoring}%
  \BibitemOpen
  \bibfield  {author} {\bibinfo {author} {\bibfnamefont {Y.}~\bibnamefont
  {Zhao}}\ and\ \bibinfo {author} {\bibfnamefont {A.}~\bibnamefont {Alu}},\
  }\href@noop {} {\bibfield  {journal} {\bibinfo  {journal} {Nano Letters}\
  }\textbf {\bibinfo {volume} {13}},\ \bibinfo {pages} {1086} (\bibinfo {year}
  {2013})}\BibitemShut {NoStop}%
\bibitem [{\citenamefont {jun Liu}\ \emph {et~al.}(2017)\citenamefont {jun
  Liu}, \citenamefont {yin Xiao},\ and\ \citenamefont {hua
  Wang}}]{jun2017multi}%
  \BibitemOpen
  \bibfield  {author} {\bibinfo {author} {\bibfnamefont {D.}~\bibnamefont {jun
  Liu}}, \bibinfo {author} {\bibfnamefont {Z.}~\bibnamefont {yin Xiao}},\ and\
  \bibinfo {author} {\bibfnamefont {Z.}~\bibnamefont {hua Wang}},\ }\href@noop
  {} {\bibfield  {journal} {\bibinfo  {journal} {Plasmonics}\ }\textbf
  {\bibinfo {volume} {12}},\ \bibinfo {pages} {445} (\bibinfo {year}
  {2017})}\BibitemShut {NoStop}%
\bibitem [{\citenamefont {Yu}\ \emph {et~al.}(2012)\citenamefont {Yu},
  \citenamefont {Aieta}, \citenamefont {Genevet}, \citenamefont {Kats},
  \citenamefont {Gaburro},\ and\ \citenamefont {Capasso}}]{yu2012broadband}%
  \BibitemOpen
  \bibfield  {author} {\bibinfo {author} {\bibfnamefont {N.}~\bibnamefont
  {Yu}}, \bibinfo {author} {\bibfnamefont {F.}~\bibnamefont {Aieta}}, \bibinfo
  {author} {\bibfnamefont {P.}~\bibnamefont {Genevet}}, \bibinfo {author}
  {\bibfnamefont {M.~A.}\ \bibnamefont {Kats}}, \bibinfo {author}
  {\bibfnamefont {Z.}~\bibnamefont {Gaburro}},\ and\ \bibinfo {author}
  {\bibfnamefont {F.}~\bibnamefont {Capasso}},\ }\href@noop {} {\bibfield
  {journal} {\bibinfo  {journal} {Nano letters}\ }\textbf {\bibinfo {volume}
  {12}},\ \bibinfo {pages} {6328} (\bibinfo {year} {2012})}\BibitemShut
  {NoStop}%
\bibitem [{\citenamefont {Yu}\ \emph {et~al.}(2011)\citenamefont {Yu},
  \citenamefont {Genevet}, \citenamefont {Kats}, \citenamefont {Aieta},
  \citenamefont {Tetienne}, \citenamefont {Capasso},\ and\ \citenamefont
  {Gaburro}}]{yu2011light}%
  \BibitemOpen
  \bibfield  {author} {\bibinfo {author} {\bibfnamefont {N.}~\bibnamefont
  {Yu}}, \bibinfo {author} {\bibfnamefont {P.}~\bibnamefont {Genevet}},
  \bibinfo {author} {\bibfnamefont {M.~A.}\ \bibnamefont {Kats}}, \bibinfo
  {author} {\bibfnamefont {F.}~\bibnamefont {Aieta}}, \bibinfo {author}
  {\bibfnamefont {J.-P.}\ \bibnamefont {Tetienne}}, \bibinfo {author}
  {\bibfnamefont {F.}~\bibnamefont {Capasso}},\ and\ \bibinfo {author}
  {\bibfnamefont {Z.}~\bibnamefont {Gaburro}},\ }\href@noop {} {\bibfield
  {journal} {\bibinfo  {journal} {science}\ }\textbf {\bibinfo {volume}
  {334}},\ \bibinfo {pages} {333} (\bibinfo {year} {2011})}\BibitemShut
  {NoStop}%
\bibitem [{\citenamefont {Kang}\ \emph {et~al.}(2012)\citenamefont {Kang},
  \citenamefont {Feng}, \citenamefont {Wang},\ and\ \citenamefont
  {Li}}]{kang2012wave}%
  \BibitemOpen
  \bibfield  {author} {\bibinfo {author} {\bibfnamefont {M.}~\bibnamefont
  {Kang}}, \bibinfo {author} {\bibfnamefont {T.}~\bibnamefont {Feng}}, \bibinfo
  {author} {\bibfnamefont {H.-T.}\ \bibnamefont {Wang}},\ and\ \bibinfo
  {author} {\bibfnamefont {J.}~\bibnamefont {Li}},\ }\href@noop {} {\bibfield
  {journal} {\bibinfo  {journal} {Optics express}\ }\textbf {\bibinfo {volume}
  {20}},\ \bibinfo {pages} {15882} (\bibinfo {year} {2012})}\BibitemShut
  {NoStop}%
\bibitem [{\citenamefont {Li}\ \emph {et~al.}(2014)\citenamefont {Li},
  \citenamefont {Guo}, \citenamefont {Wang}, \citenamefont {Zhang},
  \citenamefont {Zhang}, \citenamefont {Liu}, \citenamefont {Qu},\ and\
  \citenamefont {Gao}}]{li2014ultra}%
  \BibitemOpen
  \bibfield  {author} {\bibinfo {author} {\bibfnamefont {R.}~\bibnamefont
  {Li}}, \bibinfo {author} {\bibfnamefont {Z.}~\bibnamefont {Guo}}, \bibinfo
  {author} {\bibfnamefont {W.}~\bibnamefont {Wang}}, \bibinfo {author}
  {\bibfnamefont {J.}~\bibnamefont {Zhang}}, \bibinfo {author} {\bibfnamefont
  {A.}~\bibnamefont {Zhang}}, \bibinfo {author} {\bibfnamefont
  {J.}~\bibnamefont {Liu}}, \bibinfo {author} {\bibfnamefont {S.}~\bibnamefont
  {Qu}},\ and\ \bibinfo {author} {\bibfnamefont {J.}~\bibnamefont {Gao}},\
  }\href@noop {} {\bibfield  {journal} {\bibinfo  {journal} {Optics express}\
  }\textbf {\bibinfo {volume} {22}},\ \bibinfo {pages} {27968} (\bibinfo {year}
  {2014})}\BibitemShut {NoStop}%
\bibitem [{\citenamefont {Han}\ \emph {et~al.}(2018)\citenamefont {Han},
  \citenamefont {Ohno}, \citenamefont {Tokizane}, \citenamefont {Nawata},
  \citenamefont {Notake}, \citenamefont {Takida},\ and\ \citenamefont
  {Minamide}}]{han2018off}%
  \BibitemOpen
  \bibfield  {author} {\bibinfo {author} {\bibfnamefont {Z.}~\bibnamefont
  {Han}}, \bibinfo {author} {\bibfnamefont {S.}~\bibnamefont {Ohno}}, \bibinfo
  {author} {\bibfnamefont {Y.}~\bibnamefont {Tokizane}}, \bibinfo {author}
  {\bibfnamefont {K.}~\bibnamefont {Nawata}}, \bibinfo {author} {\bibfnamefont
  {T.}~\bibnamefont {Notake}}, \bibinfo {author} {\bibfnamefont
  {Y.}~\bibnamefont {Takida}},\ and\ \bibinfo {author} {\bibfnamefont
  {H.}~\bibnamefont {Minamide}},\ }\href@noop {} {\bibfield  {journal}
  {\bibinfo  {journal} {Optics letters}\ }\textbf {\bibinfo {volume} {43}},\
  \bibinfo {pages} {2977} (\bibinfo {year} {2018})}\BibitemShut {NoStop}%
\bibitem [{\citenamefont {Li}\ \emph {et~al.}(2019)\citenamefont {Li},
  \citenamefont {Guo}, \citenamefont {Xu}, \citenamefont {Chen}, \citenamefont
  {Hang}, \citenamefont {Zhou},\ and\ \citenamefont {Chen}}]{li2019multiple}%
  \BibitemOpen
  \bibfield  {author} {\bibinfo {author} {\bibfnamefont {J.}~\bibnamefont
  {Li}}, \bibinfo {author} {\bibfnamefont {H.}~\bibnamefont {Guo}}, \bibinfo
  {author} {\bibfnamefont {T.}~\bibnamefont {Xu}}, \bibinfo {author}
  {\bibfnamefont {L.}~\bibnamefont {Chen}}, \bibinfo {author} {\bibfnamefont
  {Z.}~\bibnamefont {Hang}}, \bibinfo {author} {\bibfnamefont {L.}~\bibnamefont
  {Zhou}},\ and\ \bibinfo {author} {\bibfnamefont {S.}~\bibnamefont {Chen}},\
  }\href@noop {} {\bibfield  {journal} {\bibinfo  {journal} {Physical Review
  Applied}\ }\textbf {\bibinfo {volume} {11}},\ \bibinfo {pages} {044042}
  (\bibinfo {year} {2019})}\BibitemShut {NoStop}%
\bibitem [{\citenamefont {Gansel}\ \emph {et~al.}(2009)\citenamefont {Gansel},
  \citenamefont {Thiel}, \citenamefont {Rill}, \citenamefont {Decker},
  \citenamefont {Bade}, \citenamefont {Saile}, \citenamefont {von Freymann},
  \citenamefont {Linden},\ and\ \citenamefont {Wegener}}]{gansel2009gold}%
  \BibitemOpen
  \bibfield  {author} {\bibinfo {author} {\bibfnamefont {J.~K.}\ \bibnamefont
  {Gansel}}, \bibinfo {author} {\bibfnamefont {M.}~\bibnamefont {Thiel}},
  \bibinfo {author} {\bibfnamefont {M.~S.}\ \bibnamefont {Rill}}, \bibinfo
  {author} {\bibfnamefont {M.}~\bibnamefont {Decker}}, \bibinfo {author}
  {\bibfnamefont {K.}~\bibnamefont {Bade}}, \bibinfo {author} {\bibfnamefont
  {V.}~\bibnamefont {Saile}}, \bibinfo {author} {\bibfnamefont
  {G.}~\bibnamefont {von Freymann}}, \bibinfo {author} {\bibfnamefont
  {S.}~\bibnamefont {Linden}},\ and\ \bibinfo {author} {\bibfnamefont
  {M.}~\bibnamefont {Wegener}},\ }\href@noop {} {\bibfield  {journal} {\bibinfo
   {journal} {Science}\ }\textbf {\bibinfo {volume} {325}},\ \bibinfo {pages}
  {1513} (\bibinfo {year} {2009})}\BibitemShut {NoStop}%
\bibitem [{\citenamefont {Shaltout}\ \emph {et~al.}(2014)\citenamefont
  {Shaltout}, \citenamefont {Liu}, \citenamefont {Shalaev},\ and\ \citenamefont
  {Kildishev}}]{shaltout2014optically}%
  \BibitemOpen
  \bibfield  {author} {\bibinfo {author} {\bibfnamefont {A.}~\bibnamefont
  {Shaltout}}, \bibinfo {author} {\bibfnamefont {J.}~\bibnamefont {Liu}},
  \bibinfo {author} {\bibfnamefont {V.~M.}\ \bibnamefont {Shalaev}},\ and\
  \bibinfo {author} {\bibfnamefont {A.~V.}\ \bibnamefont {Kildishev}},\
  }\href@noop {} {\bibfield  {journal} {\bibinfo  {journal} {Nano Letters}\
  }\textbf {\bibinfo {volume} {14}},\ \bibinfo {pages} {4426} (\bibinfo {year}
  {2014})}\BibitemShut {NoStop}%
\bibitem [{\citenamefont {Liu}\ \emph {et~al.}(2018)\citenamefont {Liu},
  \citenamefont {Zhai}, \citenamefont {Meng}, \citenamefont {Lin},
  \citenamefont {Huang}, \citenamefont {Zhao},\ and\ \citenamefont
  {Wang}}]{liu2018dirac}%
  \BibitemOpen
  \bibfield  {author} {\bibinfo {author} {\bibfnamefont {G.-D.}\ \bibnamefont
  {Liu}}, \bibinfo {author} {\bibfnamefont {X.}~\bibnamefont {Zhai}}, \bibinfo
  {author} {\bibfnamefont {H.-Y.}\ \bibnamefont {Meng}}, \bibinfo {author}
  {\bibfnamefont {Q.}~\bibnamefont {Lin}}, \bibinfo {author} {\bibfnamefont
  {Y.}~\bibnamefont {Huang}}, \bibinfo {author} {\bibfnamefont {C.-J.}\
  \bibnamefont {Zhao}},\ and\ \bibinfo {author} {\bibfnamefont {L.-L.}\
  \bibnamefont {Wang}},\ }\href@noop {} {\bibfield  {journal} {\bibinfo
  {journal} {Optics Express}\ }\textbf {\bibinfo {volume} {26}},\ \bibinfo
  {pages} {11471} (\bibinfo {year} {2018})}\BibitemShut {NoStop}%
\bibitem [{\citenamefont {Chen}\ \emph {et~al.}(2017)\citenamefont {Chen},
  \citenamefont {Zhang}, \citenamefont {Liu}, \citenamefont {Zhao},
  \citenamefont {Guo},\ and\ \citenamefont {Zhang}}]{chen2017realization}%
  \BibitemOpen
  \bibfield  {author} {\bibinfo {author} {\bibfnamefont {H.}~\bibnamefont
  {Chen}}, \bibinfo {author} {\bibfnamefont {H.}~\bibnamefont {Zhang}},
  \bibinfo {author} {\bibfnamefont {M.}~\bibnamefont {Liu}}, \bibinfo {author}
  {\bibfnamefont {Y.}~\bibnamefont {Zhao}}, \bibinfo {author} {\bibfnamefont
  {X.}~\bibnamefont {Guo}},\ and\ \bibinfo {author} {\bibfnamefont
  {Y.}~\bibnamefont {Zhang}},\ }\href@noop {} {\bibfield  {journal} {\bibinfo
  {journal} {Optical Materials Express}\ }\textbf {\bibinfo {volume} {7}},\
  \bibinfo {pages} {3397} (\bibinfo {year} {2017})}\BibitemShut {NoStop}%
\bibitem [{\citenamefont {Cai}\ \emph {et~al.}(2020)\citenamefont {Cai},
  \citenamefont {Guo}, \citenamefont {Zhou}, \citenamefont {Huang},
  \citenamefont {Yang},\ and\ \citenamefont {Zhu}}]{cai2020tunable}%
  \BibitemOpen
  \bibfield  {author} {\bibinfo {author} {\bibfnamefont {Y.}~\bibnamefont
  {Cai}}, \bibinfo {author} {\bibfnamefont {Y.}~\bibnamefont {Guo}}, \bibinfo
  {author} {\bibfnamefont {Y.}~\bibnamefont {Zhou}}, \bibinfo {author}
  {\bibfnamefont {X.}~\bibnamefont {Huang}}, \bibinfo {author} {\bibfnamefont
  {G.}~\bibnamefont {Yang}},\ and\ \bibinfo {author} {\bibfnamefont
  {J.}~\bibnamefont {Zhu}},\ }\href@noop {} {\bibfield  {journal} {\bibinfo
  {journal} {Optics express}\ }\textbf {\bibinfo {volume} {28}},\ \bibinfo
  {pages} {31524} (\bibinfo {year} {2020})}\BibitemShut {NoStop}%
\bibitem [{\citenamefont {Wang}\ \emph
  {et~al.}(2020{\natexlab{b}})\citenamefont {Wang}, \citenamefont {Zhang},
  \citenamefont {Zhang}, \citenamefont {Zhang},\ and\ \citenamefont
  {Cao}}]{wang2020tunable}%
  \BibitemOpen
  \bibfield  {author} {\bibinfo {author} {\bibfnamefont {T.}~\bibnamefont
  {Wang}}, \bibinfo {author} {\bibfnamefont {H.}~\bibnamefont {Zhang}},
  \bibinfo {author} {\bibfnamefont {Y.}~\bibnamefont {Zhang}}, \bibinfo
  {author} {\bibfnamefont {Y.}~\bibnamefont {Zhang}},\ and\ \bibinfo {author}
  {\bibfnamefont {M.}~\bibnamefont {Cao}},\ }\href@noop {} {\bibfield
  {journal} {\bibinfo  {journal} {Optics Express}\ }\textbf {\bibinfo {volume}
  {28}},\ \bibinfo {pages} {17434} (\bibinfo {year}
  {2020}{\natexlab{b}})}\BibitemShut {NoStop}%
\end{thebibliography}%

\end{document}